\documentclass[pra,aps]{revtex4}
\input epsf.tex

\newcommand{\be}{\begin{equation}}
\newcommand{\ee}{\end{equation}}
\newcommand{\ba}{\begin{eqnarray}}
\newcommand{\ea}{\end{eqnarray}}
\newcommand{\ban}{\begin{eqnarray*}}
\newcommand{\ean}{\end{eqnarray*}}

\newcommand{\braket}[2]{\mbox{$ \langle #1 | #2 \rangle $}}

\newcommand{\ket}[1]{\mbox{$ | #1 \rangle $}}
\newcommand{\bra}[1]{\mbox{$ \langle #1 | $}}

\newcommand{\demi}{\frac{1}{2}}

\newcommand{\real}{\begin{picture}(8,8)\put(0,0){R}\put(0,0){\line(0,1){7}}\end{picture}}

\newcommand{\myone}{\leavevmode\hbox{\small1\normalsize\kern-.33em1}}

\newcommand{\Tr}{\mbox{Tr}}

\begin{document}

\title{Upper Bounds for the Security of two Distributed-Phase Reference Protocols of Quantum Cryptography (Coherent-One-Way and Differential-Phase-Shift)}
\author{Cyril Branciard$^1$, Nicolas Gisin$^1$, Valerio Scarani$^2$}
\affiliation{$^1$ Group of Applied Physics, University of Geneva,
Geneva, Switzerland\\ $^2$ Centre for Quantum Technologies, National
University of Singapore, Singapore}
\date{\today}

\begin{abstract}
The Differential-Phase-Shift (DPS) and the Coherent-One-Way (COW)
are among the most practical protocols for quantum cryptography, and
are therefore the object of fast-paced experimental developments.
The assessment of their security is also a challenge for theorists:
the existing tools, that allow to prove security against the most
general attacks, do not apply to these two protocols in any
straightforward way. We present new upper bounds for their security
in the limit of large distances ($d \gtrsim 50$km with typical
values in optical fibers) by considering a large class of collective
attacks, namely those in which the adversary attaches ancillary
quantum systems to each pulse or to each pair of pulses. We
introduce also two modified versions of the COW protocol, which may
prove more robust than the original one.
\end{abstract}

\maketitle

\section{Introduction}

Over the recent years quantum cryptography evolved from nice physics
to a technology that could revolutionize the science of secrecy. The
basic idea, as formulated by Bennett and Brassard in 1984 (BB84),
was based on the use of individual qubits \cite{BB84}, quickly
``translated'' to individual photons. Given the lack of convenient
single photon sources, most experiments use instead weak laser
pulses. However, it was then realized that such sources sometimes
emit multiphoton pulses and are thus in danger of
photon-number-splitting (PNS) attacks \cite{pns}. The cheap
counter-measure to PNS attacks is to reduce further the intensity of
the weak laser pulses, but this solution leads to secret bit rates
that scale quadratically with the quantum channel transmission
coefficient, $r \propto t^2$. Hwang found an elegant way out of this
drawback, suggesting using more than one intensity \cite{hwang}.
This method, called decoy state implementation, allows one to
achieve a linear secret key rate, as for the historical single-qubit
protocols \cite{decoy}.

The BB84 protocol in all its implementations, several variations
thereof --- two-state \cite{b92}, six-state \cite{sixstates}, SARG04
\cite{sarg}, protocols using higher-dimensional systems
\cite{qudits}, etc. --- and all the corresponding entanglement-based
versions \cite{ekertbbm}, share a common feature: they all send
quantum symbols one by one. However, convenient telecom laser
sources emit either a continuous train of pulses (mode-locked
lasers), or a continuous wave (cw) that can be formatted by an
intensity modulator into trains of pulses. This observation led to
new protocols for efficient quantum key distribution (QKD) like the
{\em Differential-Phase-Shift (DPS)} \cite{ino03,dpsexp07} and the
{\em Coherent-One-Way (COW)} \cite{gis04,stu05} protocols. In both
protocols a continuous train of weak laser pulses is sent from the
sender, Alice, to the receiver, Bob. In the DPS protocol the
intensity of the pulses is constant, but the phase modulated. In the
COW protocol, the phases of all pulses are constant, but their
intensity modulated. The DPS and the COW protocols are so-called
\textit{distributed-phase-reference} protocols: the intervention of
an adversary, Eve, is monitored by measuring the coherence between
successive non-empty pulses. Both protocols are robust against PNS
attacks, because these can be detected \cite{stu05,ino05}; security
has also been studied against individual attacks \cite{stu05,wak06}
and more recently against some form of intercept-resend attacks
based on unambiguous state discrimination \cite{bra07,cur07,tsu07}.
However, security against the most general attacks is still elusive:
the tools, that have been developed in the last decade to tackle
this, cannot be applied in any straightforward way, because both
protocols move away from the symbol-per-symbol type of coding.

The purpose of this paper is to analyze the security of the COW and
the DPS protocols against a large class of collective attacks in the
long distance regime (i.e. when the transmission coefficient $t$ is
small). This study leads also to define variants of the COW
protocol, which make it more robust while keeping its simplicity.

The paper is structured as follows. In Section \ref{sec2}, we recall
the COW and DPS protocols, as well as some notions of security
bounds. In Section \ref{secbs}, we present the bound for security
against the beam-splitting attack (BSA) treated as a collective
attack. In Section \ref{sec2pa}, we study a family of attacks that
generalize the BSA by introducing errors. The basic idea is that the
adversary, Eve, attaches ancillary quantum systems to each pulse or
to each pair of pulses. For these attacks, bounds for security are
provided in the limits of large distances, typically $d\gtrsim 50$
km. These upper bounds on the secret key rates scale linearly with
$t$.

\section{Definitions and Tools}
\label{sec2}

The source, on Alice's side, produces weak coherent pulses. A
non-empty pulse is written $\ket{\alpha}$, its mean photon number
$\mu=|\alpha|^2$. The transmission coefficient of the quantum
channel connecting Alice and Bob is $t$, the efficiency of Bob's
photon counters is $\eta$; we neglect the effects of dark counts and
dead times of the detectors. Accordingly, in the absence of Eve,
when Alice sends $\ket{\alpha}$, Bob receives $\ket{\sqrt{t}\alpha}$
and has a probability $1-e^{-\mu t \eta}$ ($\approx \mu t \eta$ in
the limit $\mu t \ll 1$) to detect a photon.

\subsection{The COW and DPS codings}

In the COW protocol, each bit is coded in a sequence of one
non-empty and one empty pulse: the bit value $0$ is coded in the
sequence $\ket{\alpha}\ket{0}$, the bit value $1$ in the sequence
$\ket{0} \ket{\alpha}$. These two states are not orthogonal because
of the vacuum component, and can be unambiguously discriminated in
an optimal way by just measuring the time of arrival. This is the
very simple \textit{data line}, in which the raw key is created. The
quantum bit error rate (QBER) $Q$ is, as usual, the probability that
Bob accepts the wrong value of the bit: in physical terms, this
means that Bob has got a detection in a time slot, in which Alice
has sent an empty pulse. To estimate the loss of coherence in the
channel (and thence Eve's information), a fraction of the light is
sent into a \textit{monitoring line}, consisting of an unbalanced
interferometer. The phase between the two arms is chosen so that two
consecutive non-empty pulses sent by Alice should always interfere
constructively in one output port (and be detected with probability
$p_{D_0} > 0$) and destructively in the other one ($p_{D_1}=0$). The
departure from this ideal situation is measured by the visibility
$V=\frac{p_{D_0}-p_{D_1}}{p_{D_0}+p_{D_1}}$ of the interference
pattern observed for two consecutive non-empty pulses. Note that
there is no {\em a priori} relation between $Q$ and $V$.

In the DPS protocol, Alice produces a sequence of coherent states of
same intensity $...\ket{e^{i\varphi_{k-1}}\alpha}
\ket{e^{i\varphi_{k}}\alpha} \ket{e^{i\varphi_{k+1}}\alpha}...$
where each phase can be set at $\varphi=0$ or $\varphi=\pi$. The
bits are coded in the difference between two successive phases:
$b_k=0$ if $e^{i\varphi_{k}}=e^{i\varphi_{k+1}}$ and $b_k=1$
otherwise. These two cases can be unambiguously discriminated using
an unbalanced interferometer. The same interferometer provides the
information about the lack of coherence in the channel, used to
estimate Eve's information. Contrary to what happens for COW, the
QBER $Q$ and the visibility $V$ of the interference pattern are
tightly related in DPS through the relation $Q=\frac{1-V}{2}$.

\subsection{Three versions of COW}
\label{ss3cow}

In the \textit{original version} of COW, the pairing of the pulses
is known in advance; in addition to sending the two sequences that
code for a bit value, Alice should also send \textit{decoy
sequences} $\ket{\alpha}\ket{\alpha}$ with probability $f$ in order
to prevent a subtle form of photon-number-splitting attacks. Such
sequences do not code for a bit value: therefore, if they give rise
to a detection in the data line, this event must be eliminated in
sifting. Throughout this paper we will set $f\approx 0$: in fact,
all the bounds for security that we are going to use are valid only
in the asymptotic limit of infinitely long keys, in which case an
arbitrarily small amount of events is sufficient to produce
meaningful statistics.

Along with this original version, we introduce and study here
\textit{two modified versions} of COW, in which the pairing of the
pulses is not known \textit{a priori} by Bob, nor Eve. Alice and
Bob's devices are the same as in the original version: Alice sends a
train of empty or non-empty pulses; Bob measures the time of arrival
on his data line and checks the coherence of successive non-empty
pulses on his monitoring line. Only after the transmission, Alice
announces the pairings publicly; the bit is accepted if Bob has got
one and only one detection in the data line corresponding to that
pair of pulses. Given this, the two versions differ in the possible
choices of pulses to be paired.

In \textit{COWm1}, Alice still pairs consecutive pulses: this makes
it the closest analog to DPS. If she wants to use (almost) all the
pulses, she will still send sequences $\ket{\alpha}\ket{0}$ or
$\ket{0} \ket{\alpha}$, and sometimes introduce an unused pulse. In
\textit{COWm2}, Alice is allowed to pair any two pulses; obviously,
all the pulses are used. There is no simple version of DPS that
would be analog to COWm2, because in order to pair arbitrary pulses
in DPS, Bob should arbitrarily change the unbalance in the
interferometer \cite{footnote_DPSm2}.

Note that many other variants of COW can be imagined, as we mention
in Appendix \ref{app_modifs_COW}.

\subsection{Secret key rates}

We consider from now on that the two values $0$ and $1$ are equally
probable both in Alice's and in Bob's list; since this can be
obtained by public communication, there is no loss in generality in
this assumption. As said, the bound for security against the most
general attack by an eavesdropper has been elusive to date for both
COW and DPS. In this paper, we are concerned with specific attacks,
which of course define only {\em upper bounds} for security (i.e.,
it is guaranteed that one cannot obtain larger rates). In the family
of attacks that we consider, Eve interacts with the pulses
one-by-one or two-by-two, always with the same strategy. She is
allowed to keep her ancillae in a quantum memory, and to extract the
largest possible information out of them after Alice and Bob has run
the classical post-processing. Therefore, we will compute the bound
for security against {\em collective attacks} (as in most QKD
studies to date, we compute this bound for the asymptotic case of an
infinitely long raw key).

For collective attacks, Devetak and Winter \cite{dev05} have shown
that Eve's information on Alice's bits is bounded by the maximal
capacity of a channel Alice-Eve, in which Alice would code her bit
value $a$ in the state $\rho_{E}^{A=a}$. This quantity is called
Holevo bound, and reads \ba \chi_{AE}\,\equiv\,
\chi\left(\rho_{E}^{A=0},\rho_{E}^{A=1}\right)&=& S(\rho_E)-\demi
S(\rho_{E}^{A=0})-\demi S(\rho_{E}^{A=1})\label{holevochi} \ea where
$\rho_{E}=\demi \rho_{E}^{A=0}+\demi \rho_{E}^{A=1}$ is Eve's state
and $S$ is von Neumann entropy. A similar definition holds for Eve's
information on Bob's bits. Concerning the Alice-Bob channel, the
QBER is $\mbox{Prob}(A\neq B)=Q$; in particular, for the conditional
Shannon entropy it holds $H(A|B)=H(B|A)=h(Q)$ where $h$ is the
binary entropy. The Devetak-Winter bound reads, for the secret key
rate $r$: \ba
r&=&r_{sift}\,\big[1-h(Q)-\min(\chi_{AE},\chi_{BE})\big] \ea where
$r_{sift}$ is the sifting rate, i.e. the probability that Alice and
Bob accept a bit; we suppose that the two protocols are run with the
same repetition rate: the rates will be given "per time slot" (or
"per pulse", independently of whether the pulse is empty or not). We
work in the {\em trusted-device scenario}, in which one assumes that
Eve cannot modify the efficiency of Bob's detectors. Note that the
whole analysis can be readily adapted to the {\em untrusted-device
scenario} by replacing everywhere first $\eta\to 1$, then $t\to
t\eta$.

\section{Collective Beam-Splitting Attack ($Q=0$, $V=1$)}
\label{secbs}

The Beam-Splitting Attack (BSA) translates the fact that all the
light that is lost in the channel must be given to Eve. The attack
consists in Eve simulating the losses $1-t$ by putting a
beam-splitter just outside Alice's laboratory, and then forwarding
the remaining photons to Bob through a lossless line. Since it
simulates exactly Bob's expected optical mode, the BSA introduces no
errors (here, $Q=0$ and $V=1$) and is therefore impossible to detect
\cite{footnote_attacks_zero_errors}.

For both COW and DPS, Alice prepares a sequence of coherent states
$\bigotimes_k\ket{\alpha_k}$: each $\alpha_k$ is chosen in
$\{+\alpha,0\}$ for COW, in $\{+\alpha,-\alpha\}$ for DPS. Bob
receives the state $\bigotimes_k\ket{\alpha_k\sqrt{t}}$: Bob's
optical mode is not modified. Eve's state is
$\bigotimes_k\ket{\alpha_k\sqrt{1-t}}$; let us introduce the
notations $\alpha_E=\alpha\sqrt{1-t}$, $\mu_E=|\alpha_E|^2$ and \ba
\gamma_E&=&e^{-\mu_E}=e^{-\mu(1-t)}\,.\ea When Bob announces a
detection involving pulses $k-1$ and $k$, Eve shall extract the
highest possible information out of her systems, measured by the
Holevo quantity (\ref{holevochi}). The information available to Eve
differs for the two protocols, because of the different coding of
the bits.

In COW, the bit is $0$ when $\alpha_{k-1}=\alpha\,,\alpha_{k}=0$ and
is $1$ when $\alpha_{k-1}=0\,,\alpha_{k}=\alpha$; so, writing
$P_{\psi}$ the projector on $\ket{\psi}$, we have
$\rho_{E}^{A=0}=\rho_{E}^{B=0}=P_{+\alpha_E,0}$ and
$\rho_{E}^{A=1}=\rho_{E}^{B=1}=P_{0,+\alpha_E}$; therefore, noticing
that $|\braket{+\alpha_E,0}{0,+\alpha_E}|=\gamma_E$, we obtain \ba
\chi_{AE}=\chi_{BE}\equiv
\chi_{E}^{COW}(\mu,t)&=&h\!\left(\frac{1-\gamma_E}{2}\right)\,. \ea
Since in COW half of the pulses are empty, the secret key rate is
given by \ba r_{COW}(\mu,t)&=&\frac{1}{2}\,\left(1-e^{-\mu t
\eta}\right)\,\left[1-\chi_{E}^{COW}(\mu,t)\right]\,.
\label{rate_COW_BS} \ea Since BSA is a pulse-by-pulse attack
independent of the pairing, the analysis is unchanged for COWm1 and
COWm2.

In DPS, the bit is $0$ when $\alpha_{k-1}=\alpha_k$ and is $1$ when
$\alpha_{k-1}=-\alpha_{k}$. So, with similar notations as above, we
have $\rho_{E}^{A=0}=\rho_{E}^{B=0}=\demi
P_{+\alpha_E,+\alpha_E}+\demi P_{-\alpha_E,-\alpha_E}$ and
$\rho_{E}^{A=1}=\rho_{E}^{B=1}=\demi P_{+\alpha_E,-\alpha_E}+\demi
P_{-\alpha_E,+\alpha_E}$; therefore, noticing that
$|\braket{+\alpha_E}{-\alpha_E}|=\gamma_E^2$, we obtain \ba
\chi_{AE}=\chi_{BE}\equiv \chi_{E}^{DPS}(\mu,t)&=&2\,
h\!\left(\frac{1-\gamma_E^2}{2}\right)-h\!\left(\frac{1-\gamma_E^4}{2}\right)
\ea and the resulting secret key rate is \ba
r_{DPS}(\mu,t)&=&\left(1-e^{-\mu t
\eta}\right)\,\left[1-\chi_{E}^{DPS}(\mu,t)\right]\,.\label{rate_DPS_BS}
\ea

\begin{center}
\begin{figure}
\epsfxsize=7cm \epsfbox{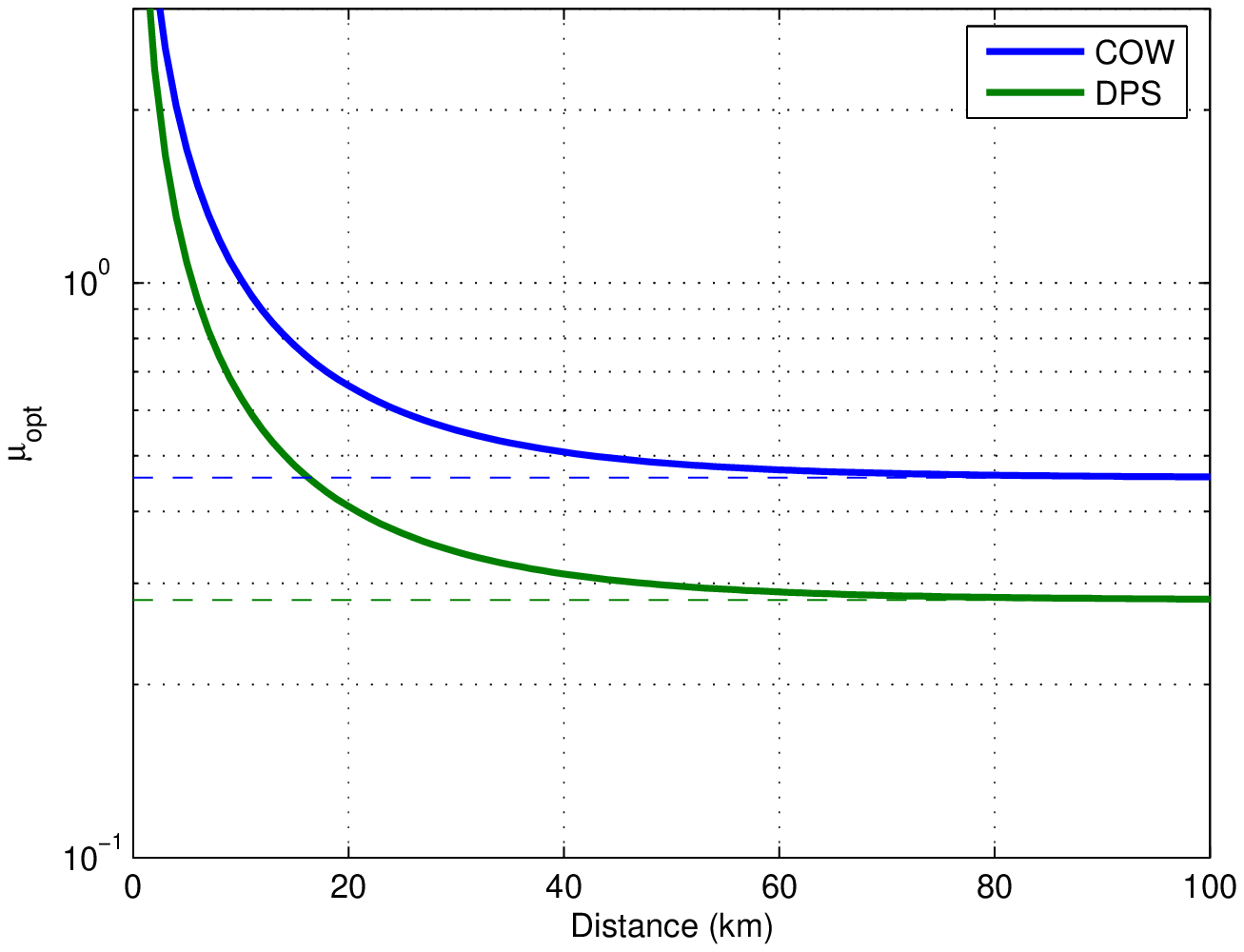} \epsfxsize=7cm \epsfbox{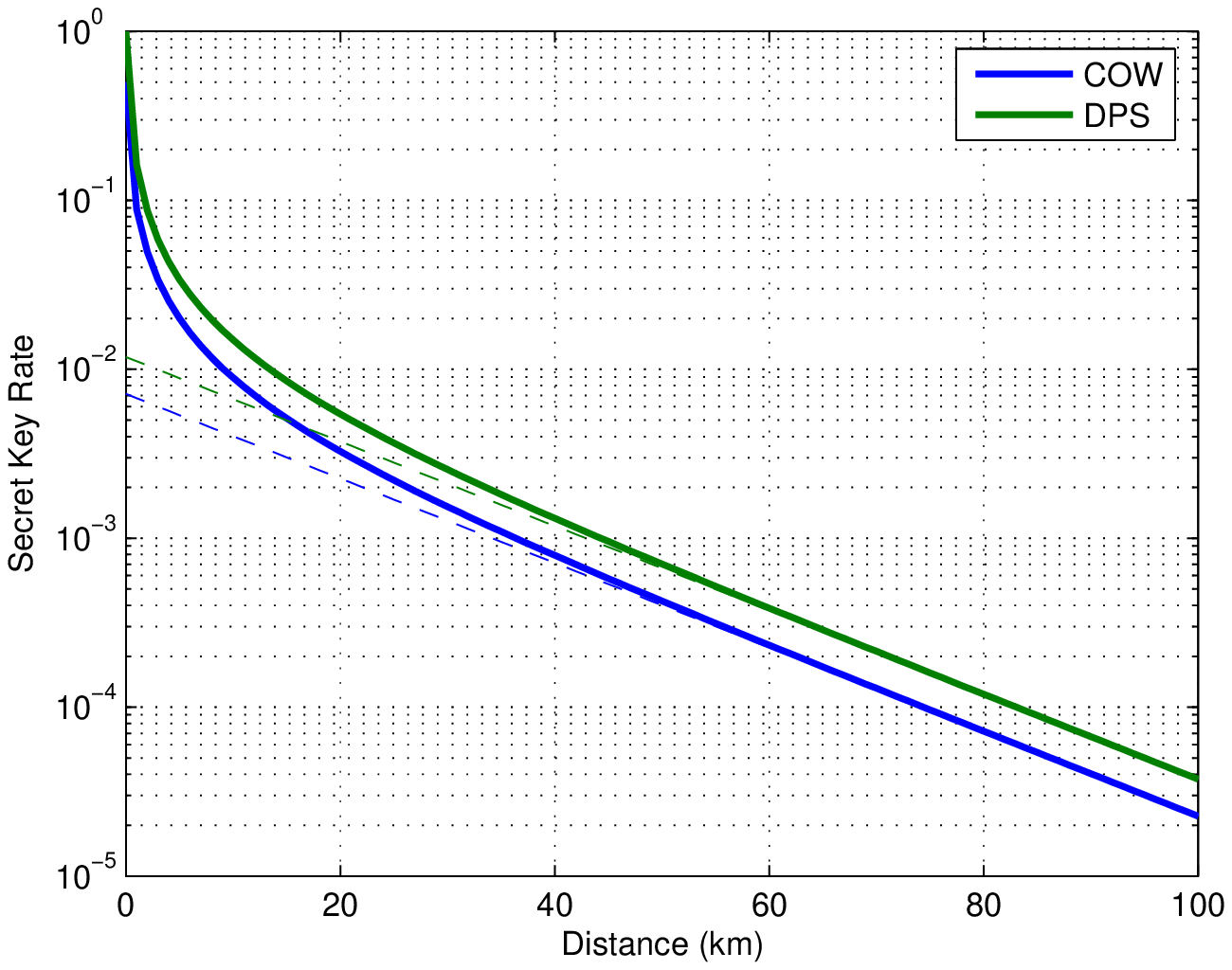}
\caption{Optimal mean photon number $\mu$ and secret key rate as a
function of the distance, for the Beam-Splitting Attack on the COW
and DPS protocols. Detection efficiency: $\eta$ = 0.1; Losses: 0.25
dB/km; no dark counts.} \label{fig_BS}
\end{figure}
\end{center}

Both for COW and DPS, Alice and Bob should choose $\mu$ such that
the secret key rate is maximized. We performed this one-parameter
optimization numerically. Figure \ref{fig_BS} shows the optimal
choice for the intensity $\mu=\mu_{opt}$ and the corresponding
secret key rates for the COW and the DPS protocols.

One notices that the two protocols show similar behaviors against
BSA. The optimal choice of $\mu$ is approximately twice as large for
COW as it is for DPS; since in COW one pulse out of two is empty,
the number of photons per bit is thus approximately the same. As for
the secret key rates obtained for the respective $\mu_{opt}$, they
are very similar, within a factor of two. The question, whether one
protocol performs better than the other one, does not have a
clear-cut answer: other practical issues should be taken into
considerations. For instance, we did not consider for COW the
fraction of the signal that should be sent through Bob's monitoring
line, which will not contribute to the key. We did not consider the
losses in Bob's interferometer neither: in DPS, they will decrease
the secret key rates, while in COW, they will not alter the key
rates. A more complete analysis should therefore lead to different
factors before the given key rates, and the factor of two that
appears here between the two protocols is not meaningful in itself.

In the limit of large distances $\mu t \ll 1$ (typically, for $d
\gtrsim 50 $km \cite{footnote_distances}), the secret key rates
under a BSA become linear in $t \eta$ ($r = r_0 t \eta$), and the
$\mu_{opt}$ tend to a constant value (dashed lines on Figure
\ref{fig_BS}). Specifically: for COW, $\mu_{opt} \to 0.4583$ and
$r_0 \approx 0.0714$; for DPS, $\mu_{opt} \to 0.2808$ and $r_0
\approx 0.1182$. Note that the attacks presented in the next Section
of this paper shall be studied only in this limit, due to their
complexity, and will coincide with the asymptotic limits for BSA
when $Q=0$ and $V=1$.

\section{Collective Attacks with $Q \geq 0$, $V \leq 1$}
\label{sec2pa}

In both COW and DPS, bits are coded in the relation between two
successive pulses. In the study of upper bounds, a natural class of
attacks is therefore the one in which Eve attacks coherently pairs
of successive pulses. These we call ``Two-Pulses Attacks'' (2PA). In
general, they are defined by \ba
[\ket{\alpha_{k-1}}\ket{\alpha_{k}}]_{A(k-1,k)}\otimes \ket{{\cal
E}}_E&\longrightarrow&
\ket{\Psi[\alpha_{k-1},\alpha_{k}]}_{B(k-1,k),E} \ea with the only
constraint that the transformation must be unitary. This class is
clearly too large to be parametrized efficiently. However, in the
limit of large distances $\mu t \ll 1$, multi-photon components on
Bob's side are supposed to be negligible; and Bob will have to
check, through the statistics of his detection rates (singles,
double-clicks in two detectors etc), that this is indeed the case.
In view of this, we restrict our study to the case where, for any
two-pulse signal sent by Alice, Bob's Hilbert space consists only of
the three orthogonal states $\ket{00}$ (no photon), $\ket{10}$ (one
photon at time $k-1$) and $\ket{01}$ (one photon at time $k$).

In this Section, 2PA are studied on COW (\ref{2pacow}), on COWm1
(\ref{2pacowmod1}) and on DPS (\ref{2padps}). On COWm2, since there
is no preferred pairing at all, we shall rather study ``One-Pulse
Attacks'', 1PA (\ref{2pacowmod2}). The resulting upper bounds will
be computed numerically and compared (\ref{numcomp}). Unless stated
otherwise, pure and mixed quantum states are normalized (in the
limit $\mu t \ll 1$) in all that follows.

\subsection{Original COW coding: Two-Pulse Attacks}
\label{2pacow}

\subsubsection{Eve's Attack}

In the original COW protocol, the pairing of the pulses sent by
Alice is publicly known. When Eve attacks the pulses two by two, we
suppose that she does it according to the same pairing. The three
sequences that Alice can send (bit 0, bit 1, decoy sequence) are
modified by Eve's intervention as \ba \begin{array}{lcl}\ket{\sqrt{\mu}, 0}_A
\ket{{\cal E}} & \quad \longrightarrow \quad & \ket{00}_B\ket{v_{\mu 0}}_E
+ \sqrt{(1-Q)\mu t}\, \ket{10}_B\ket{p_{\mu 0}^{10}}_E +
\sqrt{Q\mu t}\, \ket{01}_B\ket{p_{\mu 0}^{01}}_E \\
\ket{0, \sqrt{\mu}}_A \ket{{\cal E}} & \quad \longrightarrow \quad &
\ket{00}_B\ket{v_{0 \mu}}_E + \sqrt{Q\mu t}\, \ket{10}_B\ket{p_{0 \mu}^{10}}_E +
\sqrt{(1-Q)\mu t}\, \ket{01}_B\ket{p_{0 \mu}^{01}}_E \\
\ket{\sqrt{\mu}, \sqrt{\mu}}_A \ket{{\cal E}} & \quad
\longrightarrow \quad & \ket{00}_B\ket{v_{\mu \mu}}_E + \sqrt{(1-Q)\mu t}
\,\big( \ket{10}_B\ket{p_{\mu \mu}^{10}}_E + \ket{01}_B\ket{p_{\mu \mu}^{01}}_E
\big)
\end{array}\label{Uevecow}\ea where $\ket{v_{jk}}_E$ ($j,k \in \{0,\mu\}$) are the states that Eve attaches
to the vacuum part of the signal, while $\ket{p_{jk}^{10}}_E$ and
$\ket{p_{jk}^{01}}_E$ are the states that Eve attaches to the
1-photon part of the signal. While we have left Eve's states free
(up to some constraints to be described soon), we have fixed the
probability amplitude of each term. These amplitudes are motivated
by the expected behavior of an imperfect intensity modulator on
Alice's side, which would prepare pulses of intensity $(1-Q)\mu$ and
$Q\mu$ instead of perfectly modulated intensities $\mu$ and $0$. In
this case, for each bit sequence sent by Alice we still have an
average probability $\mu t$ that a photon arrives at Bob; in a
fraction $1-Q$ of these cases, it arrives at the correct time, in
the other cases it arrives at the wrong time, whence $Q$ is indeed
the QBER. Again, Bob has to check that the multi-photon components
are negligible.

The relations between Eve's states are constrained by the
requirement that the transformation must be unitary, and by the
results of the parameter estimation (i.e. by the values of the
visibilities). The requirement of unitarity reads (recall that we
work in the limit $\mu t \ll 1$) \ba \braket{v_{0 \mu}}{v_{\mu 0}}=
e^{-\mu}&\;,\quad& \braket{v_{\mu \mu}}{v_{\mu 0}} = \braket{v_{\mu
\mu}}{v_{0 \mu}} = e^{-\mu/2}\,.\label{unitcow}\ea The visibility in
COW is measured only conditioned to the fact that Alice has sent two
consecutive non-empty pulses. There are five such cases: the case of
decoy sequences (two non-empty pulses in the same pair) and the four
two-pair sequences $(x,y) =
(0\mu,\mu0)$,$(\mu\mu,\mu0)$,$(0\mu,\mu\mu)$,$(\mu\mu,\mu\mu)$. The
corresponding visibilities after Eve's intervention are
\ba V_{\mu\mu} &=& \mbox{Re}\big[\braket{p_{\mu\mu}^{01}}{p_{\mu\mu}^{10}}\big]\,,\label{Vd_COW}\\
V_{xy} &=&
\mbox{Re}\big[\braket{v_x}{p_x^{01}}\braket{p_y^{10}}{v_y}\big]\,.\label{Vxy_COW}\ea
As an example, consider $V_{\mu\mu}$. When Alice sends a decoy
sequence $\ket{\sqrt{\mu},\sqrt{\mu}}$, a detection in the
interferometer at the correct timing should reveal the coherence
between $\ket{10}$ and $\ket{01}$. After Eve's intervention, the
action of the interferometer (non-normalized) reads \ba
\ket{10}_B\ket{p_{\mu\mu}^{10}}_E +
\ket{01}_B\ket{p_{\mu\mu}^{01}}_E &\longrightarrow & \ket{D_0} \big(
\ket{p_{\mu\mu}^{10}}_E + \ket{p_{\mu\mu}^{01}}_E \big) + \ket{D_1}
\big( \ket{p_{\mu\mu}^{10}}_E - \ket{p_{\mu\mu}^{01}}_E \big) \,.
\label{eq_post_sel_state_COW_2} \ea The probability that the photon
going to Bob is detected by detector $D_0$ (resp. $D_1$) of the
interferometer is proportional to $p_{0/1} \propto ||
\ket{p_{\mu\mu}^{10}}_E \pm \ket{p_{\mu\mu}^{01}}_E ||^2 = 2 \pm 2
\mbox{Re} \braket{p_{\mu\mu}^{01}}{p_{\mu\mu}^{10}}\propto 1\pm
V_{\mu\mu}$, whence (\ref{Vd_COW}). The visibilities $V_{xy}$ are
computed in a similar way, considering that the interference across
the pairing is due to the coherence between $\ket{01}\ket{00}$ and
$\ket{00}\ket{10}$. In the present study, we suppose that Alice and
Bob check that all these visibilities are the same: \ba
V_{\mu\mu}\,=\,V_{0\mu,\mu0}\,=\,V_{\mu\mu,\mu0}\,=\,V_{0\mu,\mu\mu}\,=\,V_{\mu\mu,\mu\mu}&\equiv
& V\,. \label{eqvisi}\ea

\subsubsection{Eve's Information}

The task is to compute the information that Eve obtains when she
performs the attack (\ref{Uevecow}). For each bit detected by Bob,
if Eve is interested in Alice's bit, her information is the Holevo
quantity $\chi_{AE}$ computed for $\rho_E^{A = 0} = (1-Q)
\ket{p_{\mu0}^{10}}\bra{p_{\mu0}^{10}} + Q
\ket{p_{\mu0}^{01}}\bra{p_{\mu0}^{01}}$ and $\rho_E^{A = 1} = (1-Q)
\ket{p_{0\mu}^{01}}\bra{p_{0\mu}^{01}} + Q
\ket{p_{0\mu}^{10}}\bra{p_{0\mu}^{10}}$; if Eve is interested in
Bob's bit, her information is the Holevo quantity $\chi_{BE}$
computed for $\rho_E^{B = 0} = (1-Q)
\ket{p_{\mu0}^{10}}\bra{p_{\mu0}^{10}} + Q
\ket{p_{0\mu}^{10}}\bra{p_{0\mu}^{10}}$ and $\rho_E^{B = 1} = (1-Q)
\ket{p_{0\mu}^{01}}\bra{p_{0\mu}^{01}} + Q
\ket{p_{\mu0}^{01}}\bra{p_{\mu0}^{01}}$. These are formal
expressions, whose value has to be optimized under the constraints
(\ref{unitcow}) and (\ref{eqvisi}). Now, none of the constraints
(\ref{unitcow}), (\ref{Vd_COW}) and (\ref{Vxy_COW}) on Eve's states
involves $\ket{p_{0 \mu}^{10}}$ and $\ket{p_{\mu 0}^{01}}$. Eve can
therefore choose these two states freely, and the best choice is
obviously to take them orthogonal to one another and to all her
other states, in order to distinguish those cases perfectly. In this
case, $\chi_{AE} = \chi_{BE} = Q + (1-Q)
\chi(P_{p_{\mu0}^{10}},P_{p_{0\mu}^{01}})$, that we write explicitly
as \ba \chi_{COW}&=&
Q\,+\,(1-Q)\,h\left(\frac{1+|\braket{p_{\mu0}^{10}}{p_{0\mu}^{01}}|}{2}\right)\,.\label{chicow}\ea
In particular, Eve has all the information on Alice's and Bob's bit
when she introduces an error.

So finally, the Devetak-Winter bound for 2PA on COW in the limit $\mu t \ll 1$ reads \ba
r_{COW}(Q,V) = r_0 t \eta &\quad \textrm{with} \quad& r_0 =
\frac{1}{2}\, \mu\, \Big[1 - h(Q) - \chi_{COW}\Big]\,.
\label{rate2pacow}\ea Note that $r_0$ does not depend on $t \eta$:
the long-distance upper bound that we obtain is linear in $t$.

\subsection{COWm1 coding: Two-Pulse Attacks}
\label{2pacowmod1}

\subsubsection{Eve's Attack}

We now consider the first modified version of the COW protocol
(COWm1). In this version, the coding still implies pairs of
consecutive pulses, but the pairing is decided by Alice and Bob {\em
a posteriori}. Thus, during the exchange of quantum signals, Eve
does not know which pulses she should attack together: half of the
times, her 2PA will therefore be applied on pulses that are not
going to be paired to form a bit. In particular, now all four
sequences of two consecutive pulses are possible: the transformation
(\ref{Uevecow}) must be complemented with a fourth line \ba \ket{0,
0}_A \ket{{\cal E}} & \quad \longrightarrow \quad &
\ket{00}_B\ket{v_{00}}_E + \sqrt{Q\mu t}
\,\big(\ket{10}_B\ket{p_{00}^{10}}_E +
\ket{01}_B\ket{p_{00}^{01}}_E\big)\ea where the choice of
probability amplitude is dictated by the same considerations as
above. The requirement of unitarity consists of (\ref{unitcow}) and
of the three additional constraints \ba \braket{v_{00}}{v_{\mu\mu}}
= e^{-\mu}&\;,\quad& \braket{v_{00}}{v_{\mu0}} =
\braket{v_{00}}{v_{0\mu}} = e^{-\mu/2}\,.\ea The computation of the
loss of visibility is identical to the case of the original COW,
leading to (\ref{Vd_COW}) and (\ref{Vxy_COW}); as for that case, we
shall impose (\ref{eqvisi}). Note that the states $\ket{p_{0
\mu}^{10}}$, $\ket{p_{\mu 0}^{01}}$, $\ket{p_{00}^{10}}$,
$\ket{p_{00}^{01}}$ do not enter in any of the constraints, and can
therefore be chosen orthogonal to each other and to all other
states.

\subsubsection{Eve's Information}

When it comes to computing Eve's information, two cases have to be
treated separately:

\textit{Case 1:} the two pulses that code a bit have been attacked
together by Eve. In this case, the computation of Eve's information
is the same for the original COW protocol (\ref{2pacow}), so $\chi^{(2)}_{AE}=\chi^{(2)}_{BE}$ is given by (\ref{chicow}).

\textit{Case 2:} the two pulses that code a bit have not been
attacked together by Eve. To study this case, we must consider four
pulses. Writing $j,k,j',k' \in \{0,1\}$ and neglecting as usual the
two-photon terms, the transformation reads \ba \ket{j\sqrt{\mu},
k\sqrt{\mu}}_A \ket{j'\sqrt{\mu}, k'\sqrt{\mu}}_A \ket{{\cal E}{\cal
E}'} & \longrightarrow & \sqrt{(k+(-1)^kQ)\mu t}
\ket{0100}_B\ket{p_{j\mu,k\mu}^{01},v_{j'\mu,k'\mu}}_E\nonumber\\&&
+ \sqrt{(j'+(-1)^{j'}Q)\mu t}
\ket{0010}_B\ket{v_{j\mu,k\mu},p_{j'\mu,k'\mu}^{10}}_E +... \ea The
terms that we left out do not contribute, for we focus on the case
where Bob detects a photon in one of the two middle time-slots and
pairs precisely those slots. Moreover, {\em a posteriori} it is
decided that pulses $j'$ and $k$ form a bit $a$, i.e. Alice must
have used $j'=1-k=a$. Depending on the sequence sent by Alice and on
the bit detected by Bob, Eve's (unnormalized) state is thus \ba
\rho_{E_4}^{A = \{jk,j'k'\}, B = 0} = (k+(-1)^kQ)
\ket{p_{j\mu,k\mu}^{01},v_{j'\mu,k'\mu}}\bra{p_{j\mu,k\mu}^{01},v_{j'\mu,k'\mu}}
\\ \rho_{E_4}^{A = \{jk,j'k'\}, B = 1} = (j'+(-1)^{j'}Q)
\ket{v_{j\mu,k\mu},p_{j'\mu,k'\mu}^{10}}\bra{v_{j\mu,k\mu},p_{j'\mu,k'\mu}^{10}}\,.
\ea Eve's (now normalized) states conditioned on Alice's or on Bob's
bit become \ba \rho_{E_4}^{A = a} = \frac{1}{4} \sum_{j,k',b}
\rho_{E_4}^{A = \{j\bar{a},ak'\}, B = b}&\equiv& (1-Q)
\rho_{E_4}^{a,b=a} + Q \rho_{E_4}^{a,b=\bar{a}} \\ \rho_{E_4}^{B =
b} = \frac{1}{4} \sum_{j,k',a} \rho_{E_4}^{A = \{j\bar{a},ak'\}, B =
b} &\equiv & (1-Q) \rho_{E_4}^{a=b,b} + Q \rho_{E_4}^{a=\bar{b},b}
\ea where $\bar{a} = 1-a$ and $\bar{b} = 1-b$, and where \ba
\rho_{E_4}^{0,0} = \frac{1}{4} \sum_{j,k'}
\ket{p_{j\mu,\mu}^{01},v_{0,k'\mu}}\bra{p_{j\mu,\mu}^{01},v_{0,k'\mu}},
\quad \rho_{E_4}^{0,1} = \frac{1}{4} \sum_{j,k'}
\ket{v_{j\mu,\mu},p_{0,k'\mu}^{10}}\bra{v_{j\mu,\mu},p_{0,k'\mu}^{10}},
\\ \rho_{E_4}^{1,1} = \frac{1}{4} \sum_{j,k'} \ket{v_{j\mu,0},
p_{\mu,k'\mu}^{10}}\bra{v_{j\mu,0},p_{\mu,k'\mu}^{10}}, \quad
\rho_{E_4}^{1,0} = \frac{1}{4} \sum_{j,k'}
\ket{p_{j\mu,0}^{01},v_{\mu,k'\mu}}\bra{p_{j\mu,0}^{01},v_{\mu,k'\mu}}.
\ea As it happened for COW, $\rho_{E_4}^{1,0}$ and
$\rho_{E_4}^{0,1}$ are orthogonal to one another and to the other
two mixtures; therefore
$\chi^{(4)}_{AE}=\chi^{(4)}_{BE}=Q+(1-Q)\chi(\rho_{E_4}^{0,0},\rho_{E_4}^{1,1})$.

On average, each of these two cases happens with probability
$\demi$, so $\chi_{AE}=\chi_{BE}$ is given by \ba
\chi_{COWm1}&=&Q+\frac{1-Q}{2}\,\left\{h\left(\frac{1+|\braket{p_{\mu0}^{10}}{p_{0\mu}^{01}}|}{2}\right)+
\chi\left(\rho_{E_4}^{0,0},\rho_{E_4}^{1,1}\right)\right\}\,.\label{chim1}
\ea The Devetak-Winter bound for 2PA on COWm1 in the limit $\mu t
\ll 1$ reads \ba r_{COWm1}(Q,V) = r_0 t \eta &\quad \textrm{with}
\quad& r_0 = \frac{1}{2}\, \mu\, \Big[1 - h(Q) -
\chi_{COWm1}\Big]\,. \label{rate2pacowm1}\ea

\subsection{COWm2 coding: One-Pulse Attacks}
\label{2pacowmod2}

\subsubsection{Eve's Attack}

Let's now consider the second modified version of the COW protocol
(COWm2). In this version, Alice and Bob check the coherence on
successive pulses, but pair arbitrary pulses in order to define key
bits. In this case, there is no longer any natural definition of
2PA: almost always, Eve's pairing shall not correspond to the
pairing that defines a bit. Therefore, we obtain the upper bound on
COWm2 considering One-Pulse Attacks (1PA): we suppose that Eve
attaches a probe to each pulse sent by Alice, and performs the
transformation \ba
\begin{array}{lcl}\ket{0}_A \ket{{\cal E}} & \longrightarrow &
\ket{0}_B \ket{v_0}_E
+ \sqrt{Q \mu t} \ket{1}_B \ket{p_0}_E \\
\ket{\sqrt{\mu}}_A \ket{{\cal E}} & \longrightarrow & \ket{0}_B
\ket{v_\mu}_E + \sqrt{(1-Q) \mu t} \ket{1}_B
\ket{p_\mu}_E\end{array} \ea where $\ket{v_{0/\mu}}_E$ are the
states that Eve attaches to the vacuum part of the signal, while
$\ket{p_{0/\mu}}_E$ are the states that Eve attaches to the 1-photon
part of the signal. The probability amplitudes are fixed according
to the same physical considerations done for COW and COWm1. The
requirement of unitarity reads \ba \braket{v_0}{v_\mu} &=&
e^{-\mu/2}\,. \ea The loss of visibility introduced by Eve's
intervention is computed along the same lines as in \ref{2pacow}.
Suppose Alice sends a sequence $\ket{\sqrt{\mu}, \sqrt{\mu}}$: in
the limit $\mu t \ll 1$, where we neglect the 2-photon terms, Eve's
intervention leads to $\ket{00}_B \ket{v_\mu,v_\mu}_E + \sqrt{(1-Q)
\mu t} [\ket{10}_B \ket{p_\mu,v_\mu}_E + \ket{01}_B
\ket{v_\mu,p_\mu}_E]$ whence \ba V &=& \mbox{Re}\big[
\braket{v_\mu,p_\mu}{p_\mu,v_\mu}\big] =
|\braket{v_\mu}{p_\mu}|^2\,. \ea None of these constraints involves
$\ket{p_0}$, that can therefore be chosen orthogonal to all other
states of Eve.

\subsubsection{Eve's Information}

On any pair of pulses that define a bit, Eve's intervention has the
product structure \ba \begin{array}{lcl}\ket{\sqrt{\mu}, 0}_A
\ket{{\cal E}{\cal E}'} & \longrightarrow & \ket{00}_B
\ket{v_\mu,v_0}_E + \sqrt{(1-Q) \mu t} \ket{10}_B \ket{p_\mu,v_0}_E
+ \sqrt{Q \mu t} \ket{01}_B
\ket{v_\mu,p_0}_E\,, \\
\ket{0, \sqrt{\mu}}_A \ket{{\cal E}{\cal E}'} & \longrightarrow &
\ket{00}_B \ket{v_0,v_\mu}_E + \sqrt{Q \mu t} \ket{10}_B
\ket{p_0,v_\mu}_E + \sqrt{(1-Q) \mu t} \ket{01}_B
\ket{v_0,p_\mu}_E\,. \end{array}\ea For each bit detected by Bob, if
Eve is interested in Alice's bit, her information is the Holevo
quantity $\chi_{AE}$ computed for $\rho_E^{A = 0} = (1-Q)
\ket{p_\mu,v_0}\bra{p_\mu,v_0} + Q \ket{v_\mu,p_0}\bra{v_\mu,p_0}$
and $\rho_E^{A = 1} = (1-Q) \ket{v_0,p_\mu}\bra{v_0,p_\mu} + Q
\ket{p_0,v_\mu}\bra{p_0,v_\mu}$; if Eve is interested in Bob's bit,
her information is the Holevo quantity $\chi_{BE}$ computed for
$\rho_E^{B = 0} = (1-Q) \ket{p_\mu,v_0}\bra{p_\mu,v_0} + Q
\ket{p_0,v_\mu}\bra{p_0,v_\mu}$ and $\rho_E^{B = 1} = (1-Q)
\ket{v_0,p_\mu}\bra{v_0,p_\mu} + Q \ket{v_\mu,p_0}\bra{v_\mu,p_0}$.
Since $\ket{v_\mu,p_0}$ and $\ket{p_0,v_\mu}$ are orthogonal to one
another and to the other states $\ket{p_\mu,v_0}$ and
$\ket{v_0,p_\mu}$, we have $\chi_{AE} = \chi_{BE}$ given by
\ba\chi_{COWm2}&=&
Q+(1-Q)h\left(\frac{1+|\braket{v_{0}}{p_{\mu}}|^2}{2}\right)\,.\label{chim2}\ea
So finally, the Devetak-Winter bound for 1PA on COWm2 in the limit
$\mu t \ll 1$ reads \ba r_{COWm2}(Q,V) = r_0 t \eta &\quad
\textrm{with} \quad& r_0 = \frac{1}{2}\, \mu\, \Big[1 - h(Q) -
\chi_{COWm2}\Big]\,. \label{rate2pacowm2}\ea

\subsection{DPS coding: Two-Pulse Attacks}
\label{2padps}

We turn now to the DPS protocol and derive an upper bound for
security considering 2PA. The formalism is analog to the one
described for the COWm1 protocol in subsection \ref{2pacowmod1}, so
we go fast through many details. The main differences are of course
those related to the protocol: the different coding of bits, and the
link between $Q$ and $V$.

\subsubsection{Eve's Attack}

We suppose that Eve attaches her probe to two successive pulses sent
by Alice. Four two-pulse sequences are possible: with
$\sigma,\omega\in\{+,-\}$, Eve's intervention reads \ba \ket{\sigma
\sqrt{\mu}, \omega \sqrt{\mu}}_A \ket{{\cal E}} &\quad
\longrightarrow \quad &\ket{00}_B \ket{v_{\sigma \omega}}_E +
\sqrt{\mu t} \big( \sigma \ket{10}_B \ket{p_{\sigma \omega}^{10}}_E
+ \omega \ket{01}_B \ket{p_{\sigma \omega}^{01}}_E \big) \ea where
$\ket{v_{\sigma \omega}}_E$ are the states that Eve attaches to the
vacuum part of the signal, while $\ket{p_{\sigma \omega}^{10}}_E$
and $\ket{p_{\sigma \omega}^{01}}_E$ are the states that Eve
attaches to the 1-photon part of the signal (as before, Bob shall
check that the multi-photon components are negligible). The
transformation leads to the expected detection rate $\mu t \eta$ for
each pulse. The constraint of unitarity reads \ba
\braket{v_{++}}{v_{--}} & = & \braket{v_{+-}}{v_{-+}} = e^{-4\mu}\,,
\\ \braket{v_{++}}{v_{+-}} & = & \braket{v_{++}}{v_{-+}} =
\braket{v_{--}}{v_{+-}} = \braket{v_{--}}{v_{-+}} = e^{-2\mu}\,.\ea

The visibilities can now be computed for all possible sequences,
since there are no empty pulses. Formally, the expressions depend on
which sequence of pulses was sent, and on whether the two pulses
that interfere belong to a same or to different sequences according
to the pairing chosen by Eve. The resulting visibilities are \ba
V_{\sigma\omega} &=& \mbox{Re}\big[
\braket{p_{\sigma\omega}^{01}}{p_{\sigma\omega}^{10}}\big]\,,\\
V_{\sigma\omega,\sigma'\omega'} &=& \mbox{Re}\big[
\braket{v_{\sigma\omega}}{p_{\sigma\omega}^{01}}
\braket{p_{\sigma'\omega'}^{10}}{v_{\sigma'\omega'}}\big]\,. \ea
Again, Alice and Bob shall check that all these visibilities are
equal: for all $\sigma,\omega,\sigma',\omega'\in\{+,-\}$, \ba
V_{\sigma\omega} = V_{\sigma\omega,\sigma'\omega'} &\equiv& V\,. \ea

\subsubsection{Eve's Information}

As it happened for COWm1, when it comes to computing Eve's
information, two cases have to be treated separately:

\textit{Case 1:} the two pulses that contribute to the detected
event have been attacked together by Eve. The evolution in Bob's
interferometer is $\sigma \ket{10}_B \ket{p_{\sigma \omega}^{10}}_E
+ \omega \ket{01}_B \ket{p_{\sigma \omega}^{01}}_E \longrightarrow
\sum_b \ket{D_b}\ket{\psi_{\sigma,\omega,b}}_E$ with
$\ket{\psi_{\sigma,\omega,b}}= \sigma \ket{p_{\sigma\omega}^{10}} +
(-1)^b \omega \ket{p_{\sigma\omega}^{01}}$ (non-normalized). Writing
$\rho_{E_2}^{A = \{\sigma \omega\}, B =
b}=\ket{\psi_{\sigma,\omega,b}}\bra{\psi_{\sigma,\omega,b}}$, Eve's
normalized states conditioned on Alice's bit are $\rho_{E_2}^{A = 0}
= \frac{1}{8} \sum_{\sigma,b} \rho_{E_2}^{A = \{\sigma \sigma\}, B =
b}$ and $\rho_{E_2}^{A = 1} = \frac{1}{8} \sum_{\sigma,b}
\rho_{E_2}^{A = \{\sigma \bar{\sigma}\}, B = b}$ (where
$\bar{\sigma} = -\sigma$); Eve's normalized states conditioned on
Bob's bit are $\rho_{E_2}^{B = b} = \frac{1}{8} \sum_{\sigma,\omega}
\rho_{E_2}^{A = \{\sigma \omega\}, B = b}$. Eve's information for
this Case 1 is then \ba \chi^{(2)}_{AE}=\chi\left(\rho_{E_2}^{A =
0},\rho_{E_2}^{A = 1}\right) &\;,\quad&
\chi^{(2)}_{BE}=\chi\left(\rho_{E_2}^{B = 0},\rho_{E_2}^{B =
1}\right)\,. \ea

\textit{Case 2.} the two pulses that contribute to the detected
event have not been attacked together by Eve. Then we have to study
the four-pulse sequence, in which the bit has been produced by the
interference of pulses number two and three. The evolution in Bob's
interferometer is $\omega \ket{0100}_B \ket{p_{\sigma
\omega}^{01},v_{\sigma' \omega'}}_E + \sigma'\ket{0010}_B
\ket{v_{\sigma \omega},p_{\sigma' \omega'}^{10}}_E \longrightarrow
\sum_{b} \ket{D_b}\ket{\psi_{\sigma,\omega,\sigma',\omega',b}}_E$
with $\ket{\psi_{\sigma,\omega,\sigma',\omega',b}}= \omega
\ket{p_{\sigma\omega}^{01}} \ket{v_{\sigma'\omega'}} + (-1)^b
\sigma' \ket{v_{\sigma\omega}} \ket{p_{\sigma'\omega'}^{10}}$
(non-normalized). Writing $\rho_{E_4}^{A = \{\sigma \omega, \sigma'
\omega'\}, B = b}
=\ket{\psi_{\sigma,\omega,\sigma',\omega',b}}\bra{\psi_{\sigma,\omega,\sigma',\omega',b}}$,
Eve's normalized states conditioned on Alice's bit are
$\rho_{E_4}^{A = 0} = \frac{1}{32} \sum_{\sigma,\omega,\omega',b}
\rho_{E_4}^{A = \{\sigma \omega, \omega \omega'\}, B = b}$ and
$\rho_{E_4}^{A = 1} = \frac{1}{32} \sum_{\sigma,\omega,\omega',b}
\rho_{E_4}^{A = \{\sigma \omega, \bar{\omega} \omega'\}, B = b}$
(where $\bar{\omega} = -\omega$); Eve's normalized states
conditioned on Bob's bit are $\rho_{E_4}^{B = b} = \frac{1}{32}
\sum_{\sigma,\omega,\sigma',\omega'} \rho_{E_4}^{A = \{\sigma
\omega, \sigma' \omega'\}, B = b}$. Eve's information for this Case
2 is then \ba \chi^{(4)}_{AE}=\chi\left(\rho_{E_4}^{A =
0},\rho_{E_4}^{A = 1}\right) &\;,\quad&
\chi^{(4)}_{BE}=\chi\left(\rho_{E_4}^{B = 0},\rho_{E_4}^{B =
1}\right)\,. \ea

Each of the two cases happens with probability $\demi$. Therefore,
Eve's average information is \ba \chi_{AE} =
\frac{\chi^{(2)}_{AE}+\chi^{(4)}_{AE}}{2} &\;,\quad& \chi_{BE} =
\frac{\chi^{(2)}_{BE}+\chi^{(4)}_{BE}}{2}\,. \label{chidps}\ea

For the versions of COW, some of Eve's states could be immediately
chosen as being orthogonal to all the other ones; there is no such
simplification here. The Devetak-Winter bound for 2PA on DPS in the
limit $\mu t \ll 1$ reads \ba r_{DPS}(Q,V) = r_0 t \eta &\quad
\textrm{with}& \quad r_0 = \mu \left[1 - h\Big(\frac{1-V}{2}\Big) -
\min(\chi_{AE},\chi_{BE})\right]\,. \label{rate2padps}\ea

\subsection{Numerical Optimization and Comparison}
\label{numcomp}

In the previous subsections, we have derived upper bounds for the
secret key rate of COW (\ref{rate2pacow}), COWm1
(\ref{rate2pacowm1}), COWm2 (\ref{rate2pacowm2}) and DPS
(\ref{rate2padps}) in the limit $\mu t\ll 1$ of large distances. In
this limit, all these bounds scale linearly with losses:
$r=r_0t\eta$, where only the constant factor $r_0$ depends on the
protocol. Incidentally, we remind that for COWm1 and COWm2 we have
supposed that Alice makes the pairings; if Bob would make them, the
rates given above for these protocols should be divided by 2.

At this point, we want to evaluate these bounds. This involves a
double optimization: first, for a fixed value of $\mu$, one has to
find the strategy that maximizes Eve's information; then, one has to
find the value of $\mu$ that maximizes $r$ --- in our case, $r_0$.
The details, on how the optimizations over Eve's strategies were
performed, are given in Appendices \ref{app_COW_2},
\ref{app_mCOW_2}, \ref{app_COW_1} and \ref{app_DPS_2}. For COW and
COWm2, these optimizations could be performed analytically, and we
give the analytical expressions for Eve's optimal states. For COWm1,
the optimization was performed numerically, but we could find an
analytical expression for Eve's states, in which there remains only
three parameters to optimize. For DPS, only numerical optimizations
could be performed. The second optimization (over $\mu$) could only
be done numerically in all cases.

\begin{center}
\begin{figure}
\epsfxsize=7cm \epsfbox{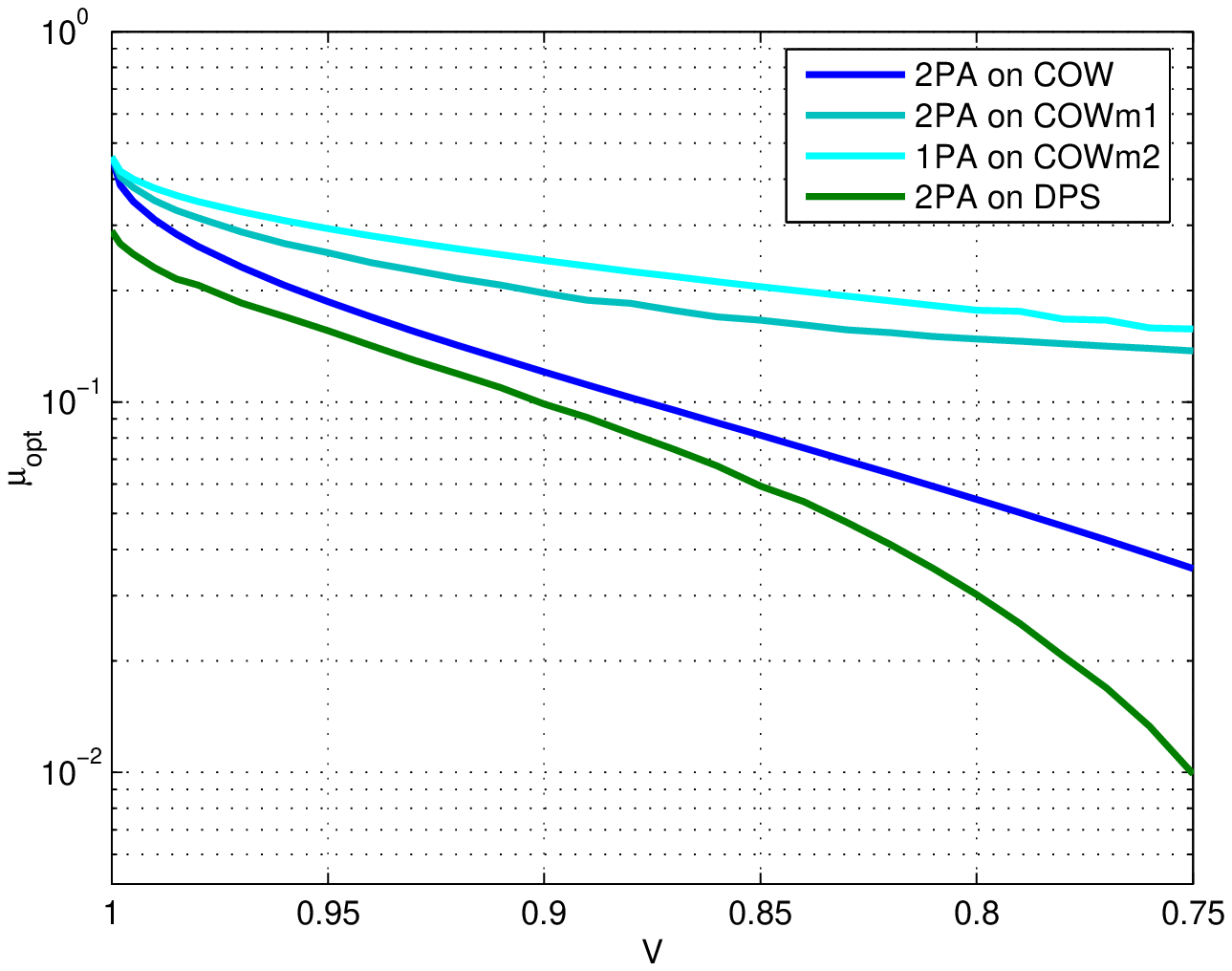} \epsfxsize=7cm
\epsfbox{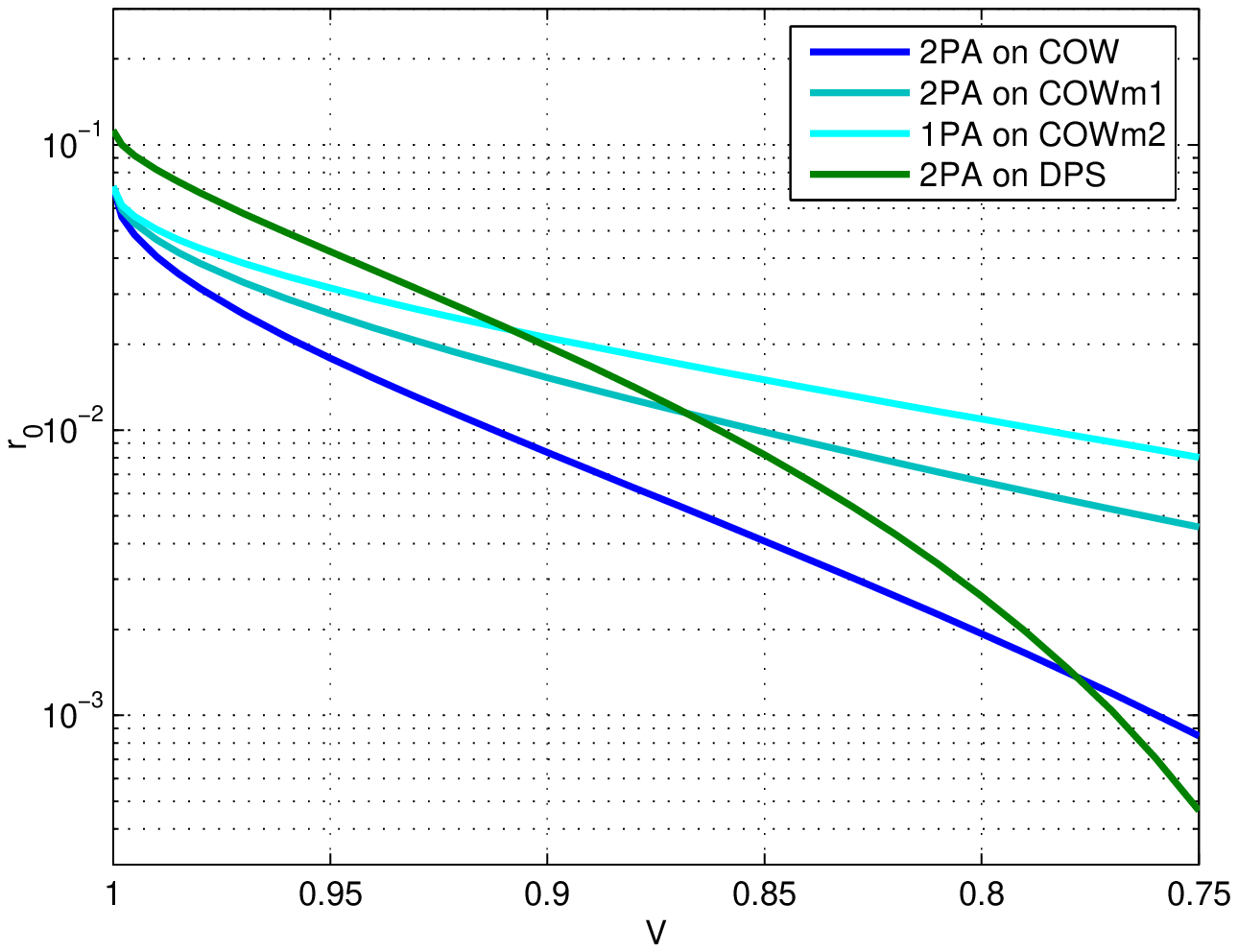} \caption{$\mu_{opt}$ and $r_0$ as a
function of the visibility $V$, for 2PA on the COW, COWm1 and DPS
protocols, and for 1PA on COWm2. For all versions of COW, we show
here the curve for $Q=0$.} \label{fig_all_Q0}
\end{figure}
\end{center}

The results of the optimizations are shown in Figure
\ref{fig_all_Q0} for the four protocols, as a function of $V$, and
in the case $Q=0$ for all versions of COW. The effect of the QBER in
the COW protocols is shown in Figures \ref{fig_COW2pa},
\ref{fig_COW2_m1} and \ref{fig_COW1}.

As expected, when $V = 1$ and $Q = 0$, the attacks under study
coincide for all protocols with the asymptotic limits for BSA. As it
was the case for BSA, one notices similar behaviors for the COW and
the DPS protocols, at least for high visibilities: the secret key
rates (or the factors $r_0$) are again very similar, within a factor
of two. Again, we cannot conclude that one protocol performs better
than the other one. The choice of which protocol to run should be
motivated by various practical reasons that we did not consider
here. Still, and as expected, the modified versions of COW provide
better bounds than the original COW: Eve's attack is less efficient
when Eve doesn't know how Alice and Bob will choose the pairing of
the pulses.

\begin{center}
\begin{figure}
\epsfxsize=7cm \epsfbox{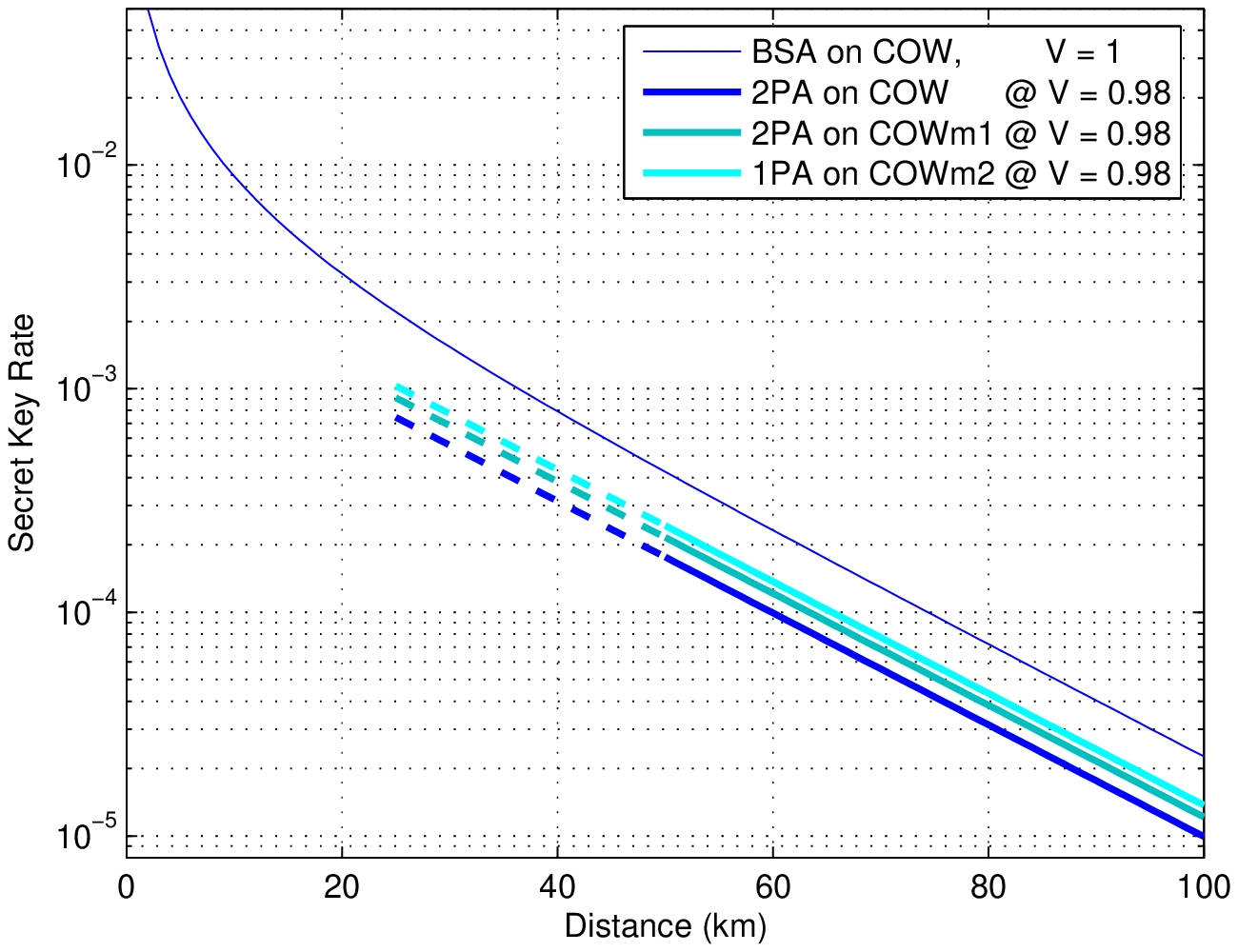} \epsfxsize=7cm
\epsfbox{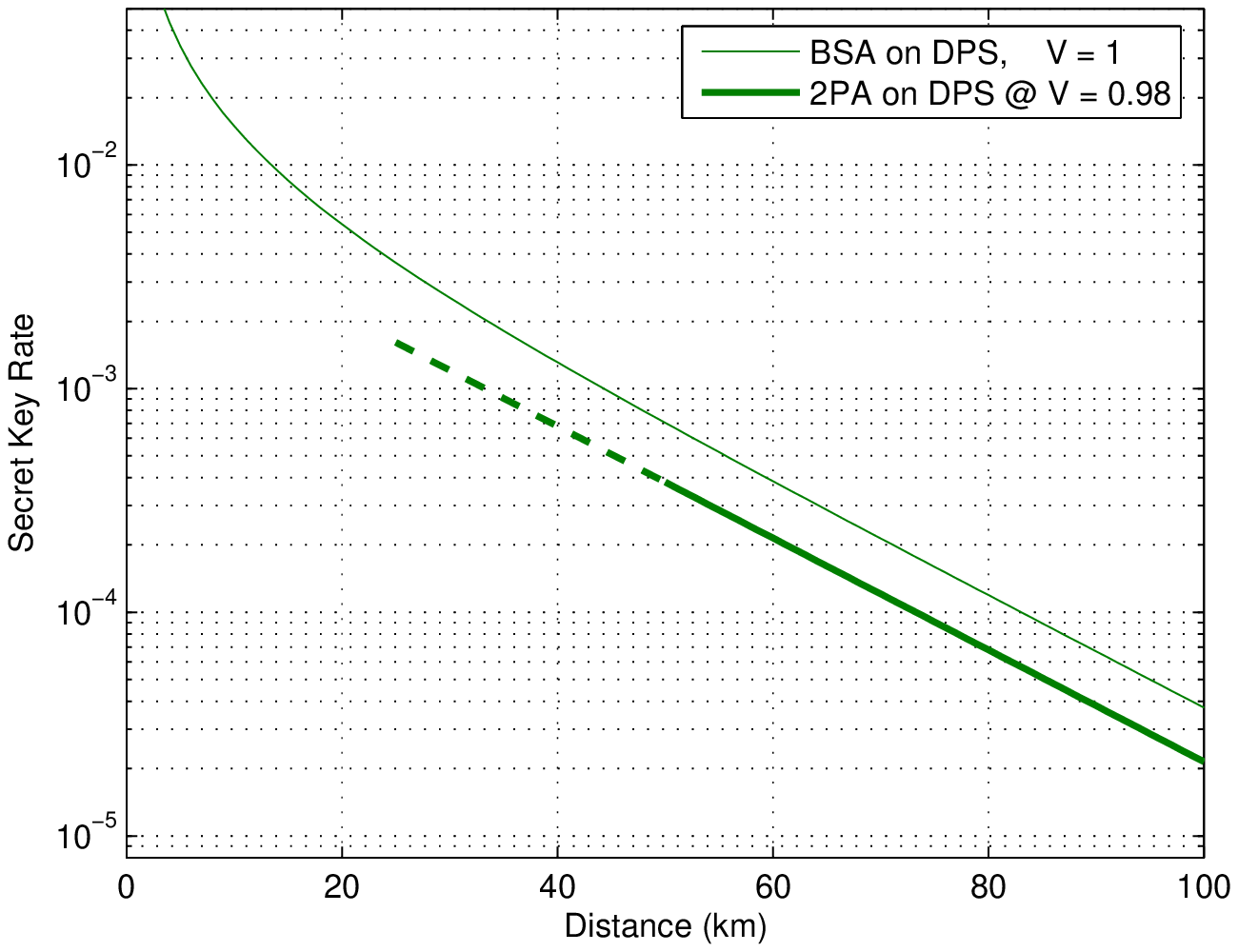} \caption{Secret key rates as a function of
the distance, for 2PA or 1PA on each protocol (for $V = 0.98$),
compared to BSA. Same parameters as in Figure \ref{fig_BS}. Recall
that the bounds obtained for 2PA and 1PA are valid only in the limit
of large distances.} \label{fig_rates_vs_dist}
\end{figure}
\end{center}

Finally, in order to get the secret key rates for a given distance,
one just has to multiply the factor $r_0$ by $t \eta$. We show as an
example in Figure \ref{fig_rates_vs_dist} the rates that we get for
each protocol in the case of $V = 0.98$ (and still $Q = 0$ for COW
and its variations), compared to BSA.

\section{Conclusion}
\label{concl}

We have provided new upper bounds for the security of the COW (the
original and two modified versions) and the DPS protocols, in the
limit of large distances. In all cases, the secret key rate goes as
$r\approx r_0 t\eta$ and scales therefore linearly with the
transmission $t$ of the channel; also, all the values of $r_0$ are
similar, within a factor of two for high visibilities. Hence, at
least given our present-day knowledge, the choice between any of
these protocols should be dictated by practical reasons rather than
by security concerns.

The two modified versions COWm1 and COWm2, introduced in a very
natural way in the context of this paper, may also prove very useful
in the future to find the bound for security against the most
general attack by the eavesdropper. Indeed, intuition suggests that
the random \textit{a posteriori} choice of the pairing may provide
the symmetry argument, which would allow to use the exponential De
Finetti theorem \cite{renner}.

\section*{Acknowledgements}

We acknowledge financial support from the European Project SECOQC
and from the Swiss NCCR ``Quantum Photonics". We are grateful to
Antonio Ac\'{\i}n, Norbert L\"utkenhaus, Renato Renner and Christoph
Simon for stimulating discussions.

\appendix

\section{Other possible variants of the COW protocol}
\label{app_modifs_COW}

In the main text, we introduced two modified versions of COW. More
generally, many other variants can be defined, as we briefly mention
in this Appendix. We give examples of such variants, that could be
useful for future studies.

There are two main motivations for looking at possible variants: one
is to find a more efficient or more robust protocol (in the present
case for instance, COWm1 and COWm2 were found to be more robust
against the attacks under study); the other one is to find a
protocol for which it should be easier to prove the security (for
instance, a protocol with more symmetries, or where the signals that
code for different bits would be more independent).

In all the following variants, Alice and Bob use the same devices as
in the original COW protocol: Alice sends a certain fraction $q$ of
weak coherent pulses $\ket{\sqrt{\mu}}$ (with an overall phase
relationship) and a fraction $1-q$ of empty pulses $\ket{0}$; Bob
measures the time of arrival on his data line and checks the
coherence of successive non-empty pulses on his monitoring line. The
only differences lie in the way the classical information is
encoded, or in how the key reconciliation is performed.

We do not provide here a security analysis for the following
variants; nonetheless, in order to give a rough idea of how the
various versions should perform, we estimate the sifting rate $r_s$
and the ideal mutual information per sifted pulse $I_{AB}$ between
Alice and Bob in the limit of large distances ($\mu t \ll 1$), in
the absence of an eavesdropper and without dark counts. The ideal
key rate would then be $r = r_s I_{AB}$.

\bigskip

The simplest possible coding is that the logical bit value 1 is
coded as a non-empty pulse $\ket{\sqrt{\mu}}$ and the bit 0 as an
empty pulse $\ket{0}$. In such a simple coding, the raw key is as
long as the entire train of pulses: $r_s=1$ (as for instance in
continuous variable QKD \cite{CV_QKD}). However, even in the absence
of an adversary and of dark counts, the error rate is large: Bob is
very likely to fail detecting a non-empty pulse, and the quantum
channel acts as a Z-channel, where the bit 0 is always detected
correctly, while the bit 1 has a high probability $e^{-\mu t \eta}$
to be detected as a 0. Straightforward application of Shannon
channel capacity shows that in the ideal case, $I_{AB} = h(q \mu t
\eta) - q h(\mu t \eta) \approx - q \log_2 q \ \mu t \eta $. For the
optimal choice of $q = e^{-1}$ one finds: $r \approx
\frac{e^{-1}}{\ln 2} \ \mu t \eta \approx 0.53 \ \mu t \eta$.

In practice, the main drawback of this basic protocol is the large
error rate. In fact, while $I_{AB}$ can in principle be extracted by
error correction (which we have supposed everywhere in this paper),
real codes do not reach this bound and become very inefficient if
the error rate is large. In other words, it is better to try and
have fewer, better correlated signals, than to keep a lot of poorly
correlated ones. One possibility to reduce this error rate is to
include a sifting step: Bob would announce his $q\mu t \eta$
fraction of time slots where he got a click on his data line, along
with another fraction $f_0$ where he had no click. In this case, the
sifting rate reduces to $r_s = q\mu t \eta + f_0$, but the fraction
of errors to be corrected is also reduced. Depending on the
practical efficiency of the error correction, one can try to
optimize $f_0$.

\bigskip

When dealing with such a Z-channel, a way to symmetrize the errors
is to code the logical bits into two physical symbols: $`0$'$ \to
\mu0$, $`1$'$ \to 0\mu$. In this prospect, the original coding of
COW appears very naturally. Contrary to the previous version, there
are no more errors due to the losses (Bob only keeps the cases where
he had one detection), and in the absence of dark counts and of Eve
$I_{AB} = 1$, and $r = r_s$.

In the original version of COW, the pairs of pulses defining each
classical bit are predefined. Alice sends pairs $\mu0$ or $0\mu$,
along with some decoy sequences $\mu\mu$ (and possibly also
sequences $00$). When the fraction of decoy sequences is negligible,
the sifting rate is $r_s = \demi \mu t \eta$.

A first possible variant of this original COW corresponds to COWm1,
where Alice still sends sequences $\mu0$ or $0\mu$, but sometimes
she introduces an unused pulse, so that the bit separations are not
known in advance by Eve. Again, if the fraction of unused pulses is
negligible, the sifting rate is $r_s = \demi \mu t \eta$.

Another variant would be that Alice sends a completely random train
of pulses $\ket{0}$ and $\ket{\sqrt{\mu}}$. She then pairs
consecutive pulses {\it a posteriori}. Here we lose a factor $\demi$
in the sifting rate ($r_s = \frac{1}{4} \mu t \eta$) because of the
sequences $00$ and $\mu\mu$ that Alice sometimes pairs together, but
the security might be easier to analyze.

In the previous two variants, one can also imagine that the pairs
are not necessarily composed of successive pulses (such as in COWm2
for instance). This might be more robust against Eve's attacks, but
this necessitates a large amount of information to be sent from
Alice to Bob for the key reconciliation.

Also, one can imagine that it is Bob who chooses the pairing: when
he gets a detection, he announces two time-slots (successive or
not), and Alice checks that they correspond to a sequence $\mu0$ or
$0\mu$. Since Bob has approximately a probability $\demi$ to
announce two time-slots that correspond to a sequence $\mu\mu$
instead, the sifting rate in this case is $r_s \approx \frac{1}{4}
\mu t \eta$.

\bigskip

Finally, one can imagine that Alice and Bob use other (longer)
sequences of pulses $\ket{\sqrt{\mu}}$ and $\ket{0}$ to encode their
classical bits (or dits). All previous variations, whether the way
the pulses are grouped is defined {\it a priori} or {\it a
posteriori}, by Alice or by Bob, whether they group successive
pulses or not, also apply to this more general variant.

\section{Optimization of 2PA on COW}
\label{app_COW_2}

We have to maximize $\chi_{COW}$ (\ref{chicow}), i.e. to minimize
$|\braket{p_{0\mu}^{01}}{p_{\mu0}^{10}}|$, submitted to the
constraints \ba \braket{v_{0 \mu}}{v_{\mu 0}}= e^{-\mu}\equiv
\gamma&\;,\quad& \braket{v_{\mu \mu}}{v_{\mu 0}} = \braket{v_{\mu
\mu}}{v_{0 \mu}} = e^{-\mu/2}\,,\label{a1cow}\ea
\ba\mbox{Re}\big[\braket{p_{\mu\mu}^{01}}{p_{\mu\mu}^{10}}\big]
\,=\,
\mbox{Re}\big[\braket{v_x}{p_x^{01}}\braket{p_y^{10}}{v_y}\big]&=& V
\label{a2cow}\ea for the four two-pair sequences $(x,y) =
(0\mu,\mu0)$,$(\mu\mu,\mu0)$,$(0\mu,\mu\mu)$,$(\mu\mu,\mu\mu)$. We
notice that the states $\ket{p_{\mu0}^{10}}$ and
$\ket{p_{0\mu}^{01}}$, whose overlap fully defines Eve's
information, are related to the states $\ket{v_{\mu0}}$ and
$\ket{v_{0\mu}}$ through (\ref{a2cow}), specifically
\ba\mbox{Re}\big[\braket{v_{0\mu}}{p_{0\mu}^{01}}\braket{p_{\mu0}^{10}}{v_{\mu0}}\big]=
V\,.\label{acow3}\ea So, we focus at first only on finding four
states $\ket{v_{\mu0}}$, $\ket{v_{0\mu}}$, $\ket{p_{\mu0}^{10}}$ and
$\ket{p_{0\mu}^{01}}$ that satisfy (\ref{acow3}) and such that
$|\braket{p_{0\mu}^{01}}{p_{\mu0}^{10}}|$ is minimal. Later, we
shall check that we can find states $\ket{v_{\mu\mu}}$,
$\ket{p_{\mu\mu}^{10}}$, and $\ket{p_{\mu\mu}^{01}}$ in order to
satisfy all the constraints (recall that the states
$\ket{p_{\mu0}^{01}}$ and $\ket{p_{0\mu}^{10}}$ are chosen to be
orthogonal to all other states).

\subsection{Parametrization of Eve's states}

First, let's choose the first 2 basis vectors such that the states
$\ket{v_{\mu0}}$ and $\ket{v_{0\mu}}$ read \ba \ket{v_{\mu0}} = \left( \begin{array}{c}
\sqrt{\frac{1+\gamma}{2}}
\\ \sqrt{\frac{1-\gamma}{2}}
\end{array} \right), \quad \ket{v_{0\mu}} = \left( \begin{array}{c} \sqrt{\frac{1+\gamma}{2}}
\\ -\sqrt{\frac{1-\gamma}{2}}
\end{array} \right)\,. \ea Let's also define $\ket{v_j^\perp}$ as the orthogonal state to
$\ket{v_j}$, in the same 2-dimensional subspace: \ba
\ket{v_{\mu0}^\perp} = \left( \begin{array}{c}
\sqrt{\frac{1-\gamma}{2}}
\\ -\sqrt{\frac{1+\gamma}{2}}
\end{array} \right), \quad \ket{v_{0\mu}^\perp} = \left( \begin{array}{c} \sqrt{\frac{1-\gamma}{2}}
\\ \sqrt{\frac{1+\gamma}{2}}
\end{array} \right)\,. \ea
We must have (\ref{acow3}). Now, if
$\braket{v_{0\mu}}{p_{0\mu}^{01}}\braket{p_{\mu0}^{10}}{v_{\mu0}}
\notin \real$, then Eve could just add a global phase to
$\ket{p_{\mu0}^{10}}$ for instance, and increase $V$ without
changing her information. This implies that Eve's maximal
information compatible with $V$ is obtained when the above quantity
is real. Then we can write, for some factor $\lambda \in [V, 1/V]$
and some phase $\tilde{\phi} \in \real$:
$\braket{v_{\mu0}}{p_{\mu0}^{10}} = \sqrt{\lambda V} e^{i
\tilde{\phi}}$ and $\braket{v_{0\mu}}{p_{0\mu}^{01}} = \sqrt{V /
\lambda} e^{i \tilde{\phi}}$. But since the phase $\tilde{\phi}$
does not play any role in Eve's information (which depends only on
$|\braket{p_{0\mu}^{01}}{p_{\mu0}^{10}}|$), we can without loss of
generality set it to 0. In conclusion, $\ket{p_{\mu0}^{10}}$ and
$\ket{p_{0\mu}^{01}}$ are of the form \ba \ket{p_{\mu0}^{10}} &=&
\sqrt{\lambda V} \ket{v_{\mu0}} - \sqrt{1-\lambda V} \cos\theta_0
e^{i \phi_0} \ket{v_{\mu0}^\perp} + \sqrt{1-\lambda V} \sin\theta_0
\ket{w_0} \\ \ket{p_{0\mu}^{01}} &=& \sqrt{V / \lambda}
\ket{v_{0\mu}} - \sqrt{1-V / \lambda} \cos\theta_1 e^{i \phi_1}
\ket{v_{0\mu}^\perp} + \sqrt{1-V / \lambda} \sin\theta_1 \ket{w_1}
\ea where $\ket{w_0}$ and $\ket{w_1}$ are any states orthogonal to
both $\ket{v_{\mu0}}$ and $\ket{v_{0\mu}}$ and $\theta_0, \theta_1,
\phi_0$ and $\phi_1$ are free parameters.

\subsection{Results of the optimization}

\begin{center}
\begin{figure}
\epsfxsize=7cm \epsfbox{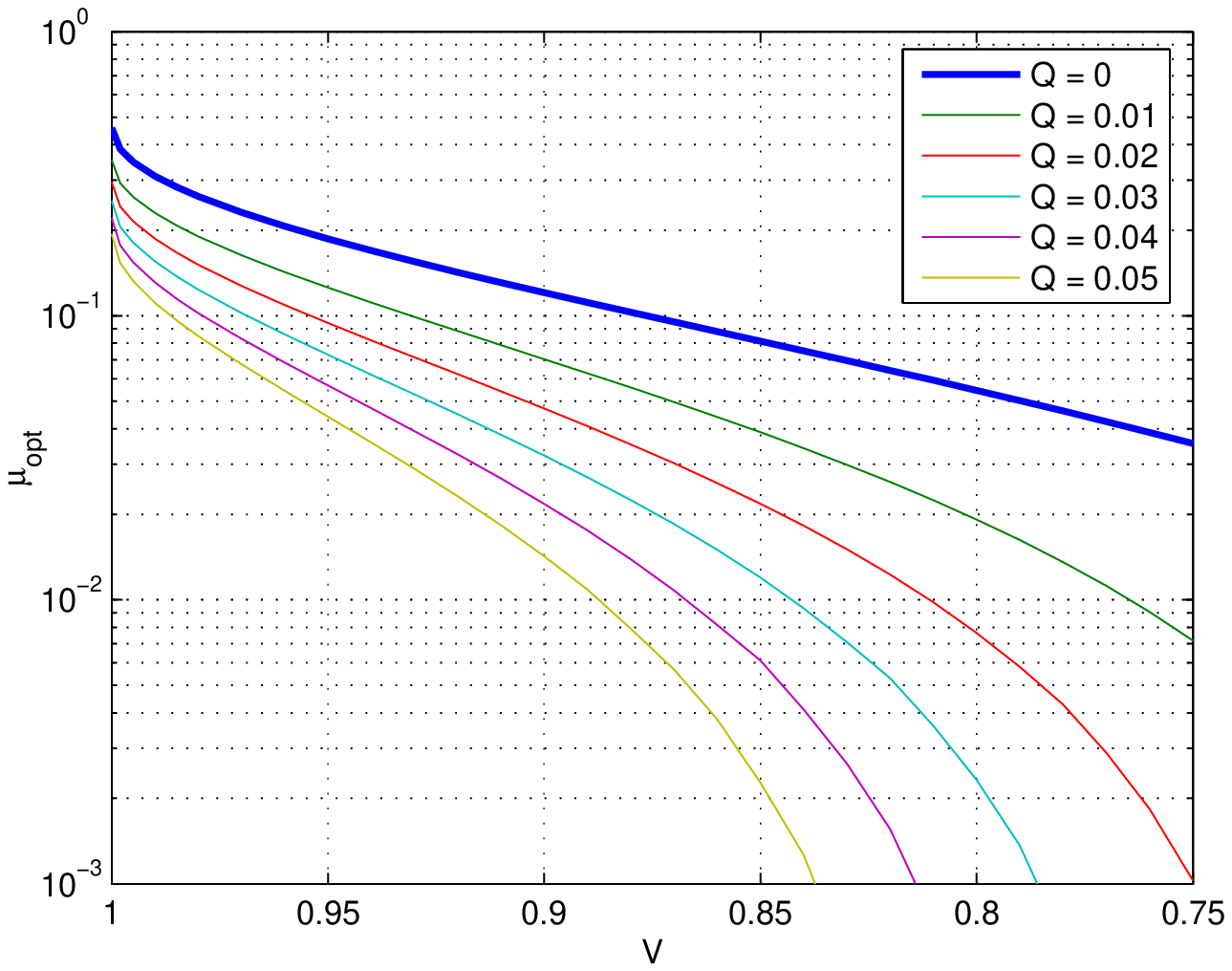} \epsfxsize=7cm
\epsfbox{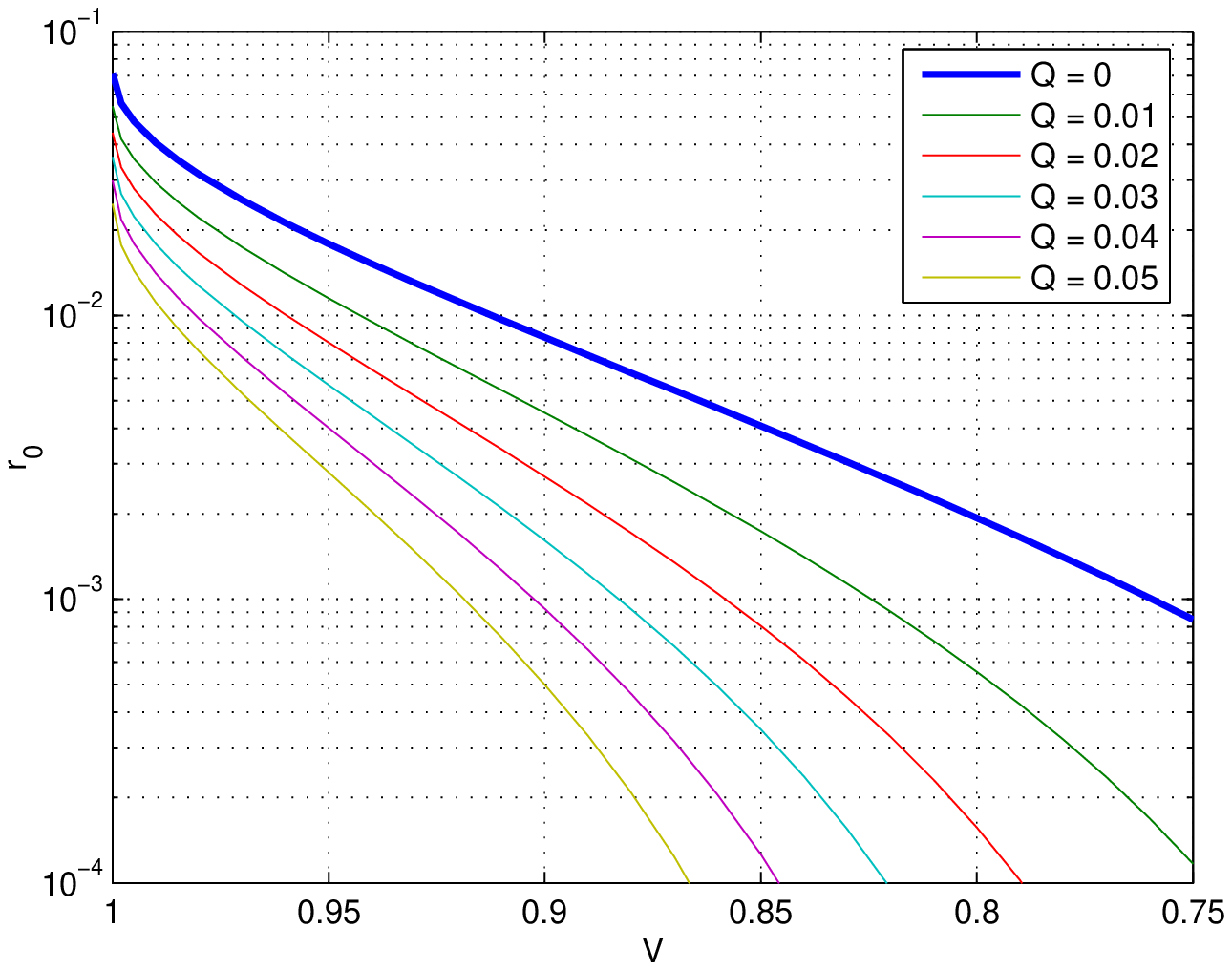} \caption{COW, original version: $\mu_{opt}$
and $r_0$ for 2PA, for different values of $Q$.} \label{fig_COW2pa}
\end{figure}
\end{center}

For $\gamma > 2 \sqrt{V(1-V)}$ (i.e. $\mu$ small enough) and $V >
1/2$, it can be proved analytically \cite{footnote_steps_proof} that
the minimum of $|\braket{p_{0\mu}^{01}}{p_{\mu0}^{10}}|$ is \ba
|\braket{p_{0\mu}^{01}}{p_{\mu0}^{10}}| & = & (2V-1)\gamma -
2\sqrt{V(1-V)}\sqrt{1-\gamma^2} \ea obtained for $\lambda = 1,
\theta_0 = \theta_1 = \phi_0 = \phi_1 = 0$, in which case
$\ket{p_{\mu0}^{10}} = \sqrt{V} \ket{v_{\mu0}} - \sqrt{1-V}
\ket{v_{\mu0}^\perp}$, $\ket{p_{0\mu}^{01}} = \sqrt{V}
\ket{v_{0\mu}} - \sqrt{1-V} \ket{v_{0\mu}^\perp}$. Having maximized
Eve's information, one can run the one-parameter optimization over
the pulse intensity $\mu$. The optimal choice $\mu_{opt}$ and the
corresponding value of $r_0$ are plotted in Fig.~\ref{fig_COW2pa},
as a function of $V$ and for different values of $Q$.

We still have to check that we can find states $\ket{v_{\mu\mu}}$,
$\ket{p_{\mu\mu}^{10}}$, and $\ket{p_{\mu\mu}^{01}}$ that satisfy
all the constraints. This is indeed the case. For instance, we complete the previous basis with a third orthogonal
vector and choose
\ba \ket{v_{\mu\mu}} = \left( \begin{array}{c}
\sqrt{\frac{2\gamma}{1+\gamma}}
\\ 0 \\ \sqrt{\frac{1-\gamma}{1+\gamma}}
\end{array} \right), \quad \ket{p_{\mu\mu}^{10}} = \left( \begin{array}{c}
\sqrt{\frac{1+V}{2}}c \\
\sqrt{\frac{1-V}{2}} \\
\sqrt{\frac{1+V}{2}}s
\end{array} \right), \quad
 \ket{p_{\mu\mu}^{01}} = \left( \begin{array}{c} \sqrt{\frac{1+V}{2}}c \\
-\sqrt{\frac{1-V}{2}} \\
\sqrt{\frac{1+V}{2}}s
\end{array} \right) \ea
with $c = \sqrt{\frac{2V}{1+V}} \sqrt{\frac{2\gamma}{1+\gamma}} +
\sqrt{\frac{1-V}{1+V}} \sqrt{\frac{1-\gamma}{1+\gamma}}$, $s =
\sqrt{\frac{2V}{1+V}} \sqrt{\frac{1-\gamma}{1+\gamma}} -
\sqrt{\frac{1-V}{1+V}} \sqrt{\frac{2\gamma}{1+\gamma}}$.
The fact, that the minimum of $|\braket{p_{0\mu}^{01}}{p_{\mu0}^{10}}|$ can be reached without using the constraints that involve the sequence $(\mu,\mu)$ means that the presence of decoy sequences does not increase the
security of COW against 2PA.

Note finally that if $\gamma \leq 2 \sqrt{V(1-V)}$ or if $V \leq
1/2$, Eve can choose her states $\ket{p_{\mu0}^{10}}$ and
$\ket{p_{0\mu}^{01}}$ (for instance $\lambda = 1, \cos\theta_0 =
\frac{\gamma V}{(1-V)\gamma + 2\sqrt{V(1-V)}\sqrt{1-\gamma^2}},
\theta_1 = \phi_0 = \phi_1 = 0$) such that
$\braket{p_{0\mu}^{01}}{p_{\mu0}^{10}} = 0$, in which case Eve can
perfectly discriminate the two states: she has the full information
on Alice and Bob's bit. Therefore, $\gamma > 2 \sqrt{V(1-V)}$ and $V
> 1/2$ are necessary conditions for Alice and Bob to establish a
secret key.

\section{Optimization of 2PA on COW$\mathrm{\bf m}$1}
\label{app_mCOW_2}

We have to maximize $\chi_{COWm1}$ (\ref{chim1}) submitted to the
constraints (\ref{a1cow}), (\ref{a2cow}) and \ba
\braket{v_{00}}{v_{\mu\mu}} = e^{-\mu}\equiv\gamma &\;,\quad&
\braket{v_{00}}{v_{\mu0}} = \braket{v_{00}}{v_{0\mu}} =
e^{-\mu/2}\,. \label{b1cow}\ea

\subsection{Parametrization of Eve's states}

We write the states $\ket{v_{jk}}$ as \ba \ket{v_{\mu0}} = \left(
\begin{array}{c} \sqrt{\frac{1+\gamma}{2}}
\\ \sqrt{\frac{1-\gamma}{2}} \\ 0 \\ 0
\end{array} \right), \quad \ket{v_{0\mu}} = \left( \begin{array}{c}
\sqrt{\frac{1+\gamma}{2}}
\\ -\sqrt{\frac{1-\gamma}{2}} \\ 0 \\ 0
\end{array} \right), \quad \ket{v_{00}} = \left( \begin{array}{c}
\sqrt{\frac{2\gamma}{1+\gamma}}
\\ 0 \\ \sqrt{\frac{1-\gamma}{1+\gamma}} \\ 0
\end{array} \right), \quad \ket{v_{\mu\mu}} = \left( \begin{array}{c}
\sqrt{\frac{2\gamma}{1+\gamma}}
\\ 0 \\ -\gamma \sqrt{\frac{1-\gamma}{1+\gamma}} \\ 1-\gamma
\end{array} \right)\,. \ea
These states satisfy the constraints (\ref{a1cow}) and
(\ref{b1cow}). We still have four states to consider,
$\ket{p_{\mu0}^{10}}, \ket{p_{0\mu}^{01}}, \ket{p_{\mu\mu}^{10}}$
and $\ket{p_{\mu\mu}^{01}}$ (recall that the states
$\ket{p_{\mu0}^{01}}, \ket{p_{0\mu}^{10}}, \ket{p_{\mu\mu}^{10}}$
and $\ket{p_{\mu\mu}^{01}}$ have already been chosen orthogonal to
all other states). Therefore, Eve's states under consideration live
in general in an eight-dimensional space. We have performed the
numerical optimization over the most general choice of the four
$\ket{p}$ states that satisfied the constraints (\ref{a2cow}).

\subsection{Results of the optimization}

The best attack we found can be realized by four-dimensional states
and depends only on three free parameters ($\theta_0, \theta_1,
\phi$), that are still to be optimized. Let's introduce the
following vectors: \ba &\ket{v_{\mu0}^\perp} = \left(
\begin{array}{c} \sqrt{\frac{1-\gamma}{2}}
\\ -\sqrt{\frac{1+\gamma}{2}} \\ 0 \\ 0
\end{array} \right), \quad \ket{v_{0\mu}^\perp} = \left( \begin{array}{c}
\sqrt{\frac{1-\gamma}{2}}
\\ \sqrt{\frac{1+\gamma}{2}} \\ 0 \\ 0
\end{array} \right), \quad \ket{v_{\mu\mu}^{\perp,1}} = \left( \begin{array}{c}
0 \\ 0 \\
\sqrt{1-\gamma^2}
\\ \gamma
\end{array} \right), \quad \ket{v_{\mu\mu}^{\perp,2}} = \left( \begin{array}{c}
\sqrt{\frac{1-\gamma}{1+\gamma}} \\ 0 \\
\gamma \sqrt{\frac{2\gamma}{1+\gamma}}
\\ -\sqrt{2\gamma}\sqrt{1-\gamma}
\end{array} \right), &\\& \quad \ket{w_2} = \left( \begin{array}{c}
0
\\ 1 \\ 0 \\ 0
\end{array} \right), \quad \ket{w_3} = \left( \begin{array}{c}
0
\\ 0 \\ 1 \\ 0
\end{array} \right), \quad \ket{w_4} = \left( \begin{array}{c}
0
\\ 0 \\ 0 \\ 1
\end{array} \right)\,. &\ea Our best attack is then defined by
\ba \ket{p_{\mu0}^{10}} = \sqrt{V} \ket{v_{\mu0}} - \sqrt{1-V}
\ket{w_{\mu0}} \, , & \quad & \ket{p_{0\mu}^{01}} = \sqrt{V}
\ket{v_{0\mu}} - \sqrt{1-V} \ket{w_{0\mu}} \, , \\
\ket{p_{\mu\mu}^{10}} = \sqrt{V} \ket{v_{\mu\mu}} - \sqrt{1-V}
\ket{w_{\mu\mu}^{10}} \, , & \quad & \ket{p_{\mu\mu}^{01}} =
\sqrt{V} \ket{v_{\mu\mu}} - \sqrt{1-V} \ket{w_{\mu\mu}^{01}} \, ,
\ea where \ba \ket{w_{\mu0}} &=& \cos\theta_0 \,
\ket{v_{\mu0}^\perp}
+ \sin\theta_0 \cos\theta_1 \, \ket{w_3}+ \sin\theta_0 \sin\theta_1 \, \ket{w_4} \, , \\
\ket{w_{0\mu}} &=& \cos\theta_0 \, \ket{v_{0\mu}^\perp}
+ \sin\theta_0 \cos\theta_1 \, \ket{w_3}+ \sin\theta_0 \sin\theta_1 \, \ket{w_4} \, , \\
\ket{w_{\mu\mu}^{10}} &=& \frac{1}{\sqrt{2}} (\cos\phi \,
\ket{v_{\mu\mu}^{\perp,1}} + \sin\phi \,
\ket{v_{\mu\mu}^{\perp,2}}) - \frac{1}{\sqrt{2}} \, \ket{w_2} \, , \\
\ket{w_{\mu\mu}^{01}} &=& \frac{1}{\sqrt{2}} (\cos\phi \,
\ket{v_{\mu\mu}^{\perp,1}} + \sin\phi \, \ket{v_{\mu\mu}^{\perp,2}})
+ \frac{1}{\sqrt{2}} \, \ket{w_2} \, . \ea Note that these states
satisfy a more constraining version of (\ref{a2cow}):
$\braket{p_{\mu\mu}^{01}}{p_{\mu\mu}^{10}} = V,
\braket{v_{0\mu}}{p_{0\mu}^{01}} =
\braket{v_{\mu\mu}}{p_{\mu\mu}^{01}} =
\braket{p_{\mu0}^{10}}{v_{\mu0}} =
\braket{p_{\mu\mu}^{10}}{v_{\mu\mu}} = \sqrt{V} $. Finally, the
optimization over the three remaining parameters $\theta_0,
\theta_1$ and $\phi$ was performed numerically.

\begin{center}
\begin{figure}
\epsfxsize=7cm \epsfbox{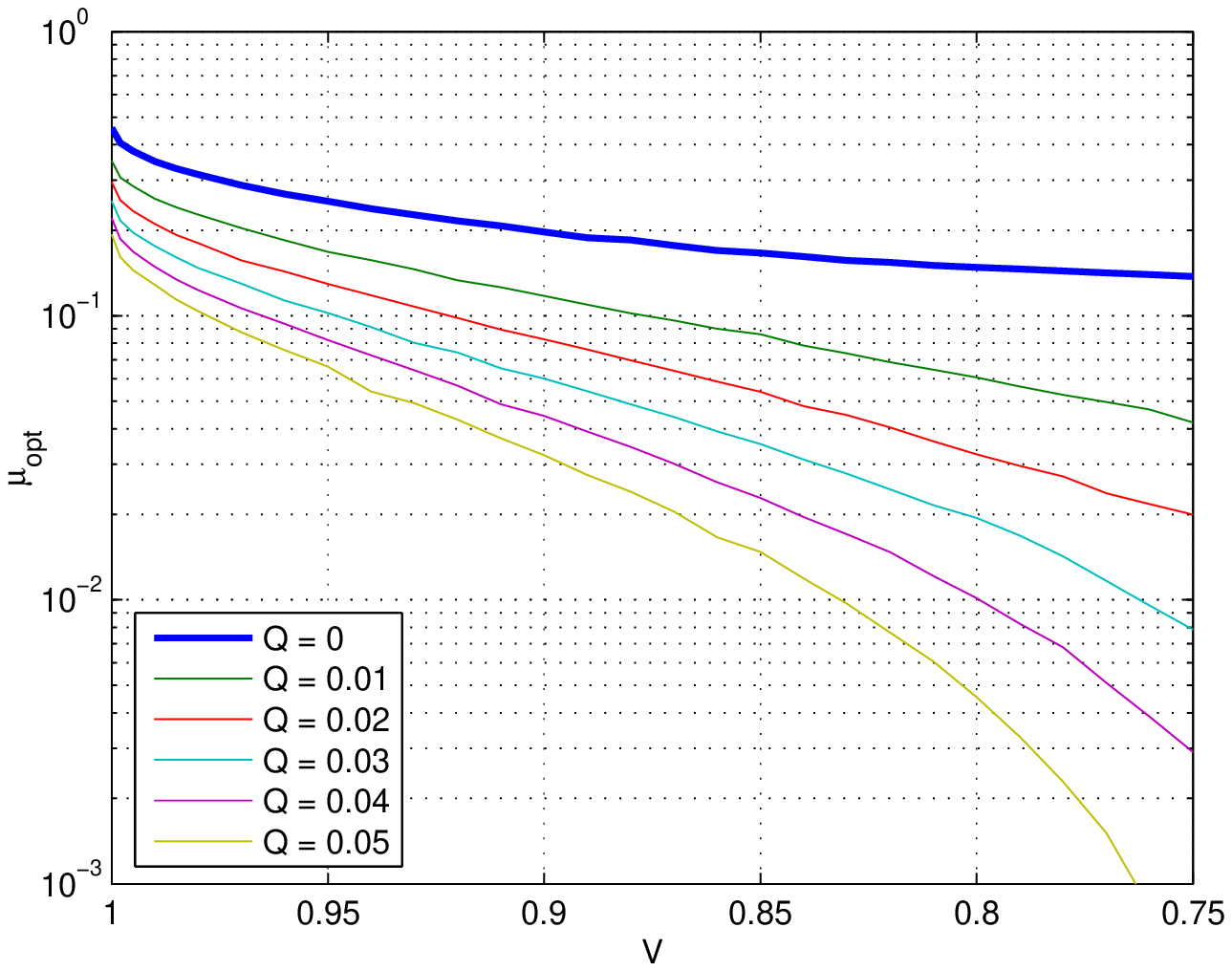} \epsfxsize=7cm
\epsfbox{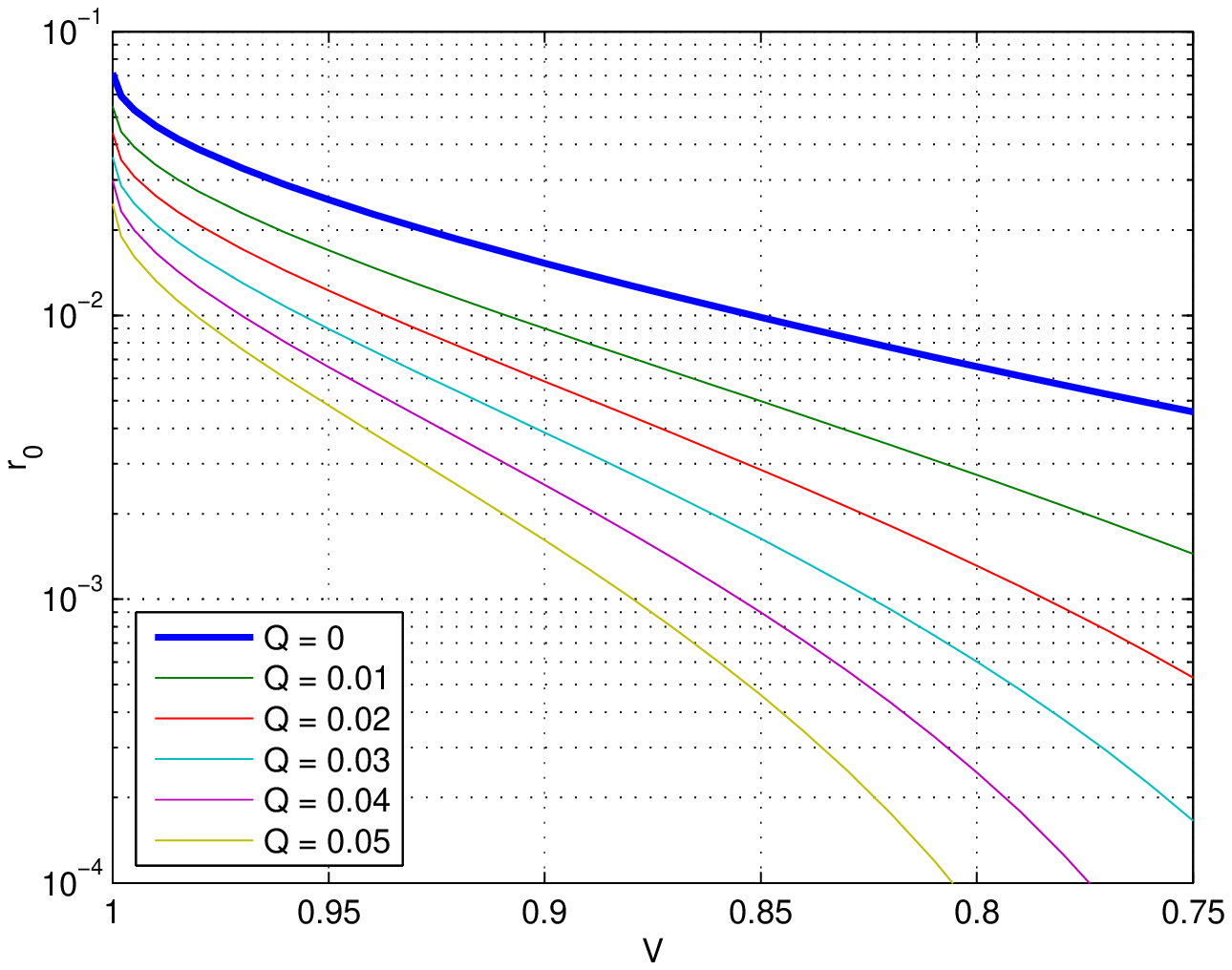} \caption{COWm1: $\mu_{opt}$ and $r_0$ for
2PA, for different values of $Q$.} \label{fig_COW2_m1}
\end{figure}
\end{center}

Having maximized Eve's information, one can run the one-parameter
optimization over the pulse intensity $\mu$. The optimal choice
$\mu_{opt}$ and the corresponding value of $r_0$ are plotted in
Fig.~\ref{fig_COW2_m1}.

\section{Optimization of 1PA on COW$\mathrm{\bf m}$2}
\label{app_COW_1}

We have to maximize $\chi_{COWm2}$ (\ref{chim2}), i.e. to minimize
$|\braket{v_{0}}{p_{\mu}}|$, submitted to the constraints \ba
\braket{v_0}{v_\mu} &=& e^{-\mu/2}\,, \\ |\braket{v_\mu}{p_\mu}|^2
&=& V\,. \ea The state $\ket{p_0}$ was already chosen to be
orthogonal to the three other states; we have therefore to work in a
three-dimensional Hilbert space. Without loss of generality, we
choose the following parametrization, which ensures automatically
that the constraints are satisfied: \ba \ket{v_\mu} = \left(
\begin{array}{c} 1 \\ 0 \\ 0
\end{array} \right), \quad \ket{v_0} = \left( \begin{array}{c} e^{-\mu/2}
\\ \sqrt{1-e^{-\mu}} \\ 0
\end{array} \right), \quad \ket{p_\mu} = e^{i \tilde{\phi}} \left( \begin{array}{c}
\sqrt{V}
\\ -\sqrt{1-V} \cos\theta e^{i \phi} \\ \sqrt{1-V} \sin\theta
\end{array} \right)\,; \ea
actually the phase $\tilde{\phi}$ does not play any role, and we set it to
be 0. So, for a given $V$ and a given
$\mu$, Eve's states are parametrized by $\theta$ and $\phi$.

For $e^{-\mu} \leq 1-V$, Eve can choose $\phi = 0$ and $\cos\theta =
\sqrt{\frac{e^{-\mu}}{1-e^{-\mu}} \frac{V}{1-V}}$, which gives
$\braket{v_0}{p_\mu} = 0$: in this case, Eve has full information on
Alice and Bob's bit. A necessary condition for Alice and Bob to have
secret bits is therefore to choose $\mu$ such that $e^{-\mu} > 1-V$.
In this case, one can easily show that the minimum overlap is \ba
|\braket{v_0}{p_\mu}| &=& e^{-\mu/2}\sqrt{V} - \sqrt{1-e^{-\mu}}
\sqrt{1-V}\ea obtained by setting $\theta = \phi = 0$.

\begin{center}
\begin{figure}
\epsfxsize=7cm \epsfbox{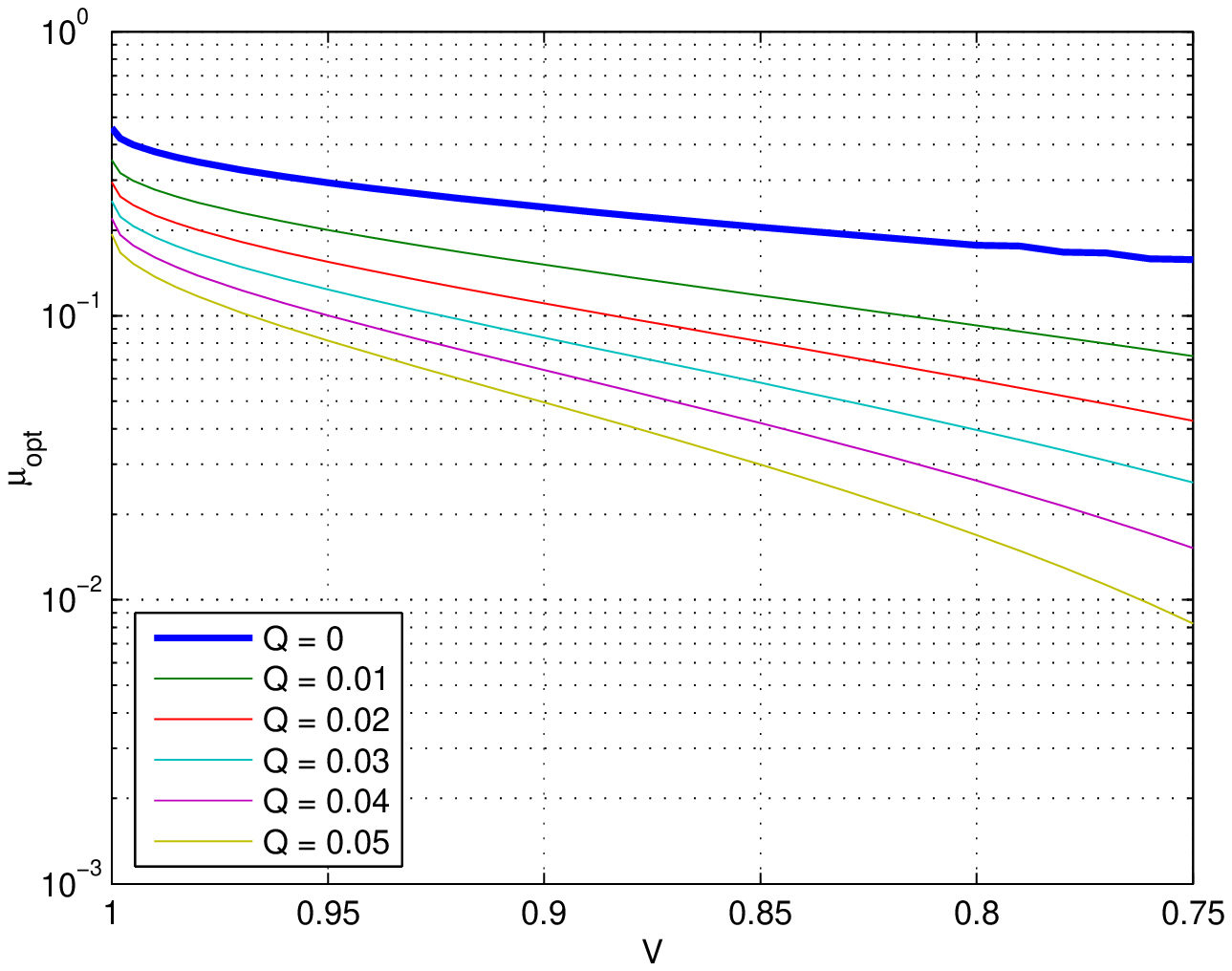} \epsfxsize=7cm
\epsfbox{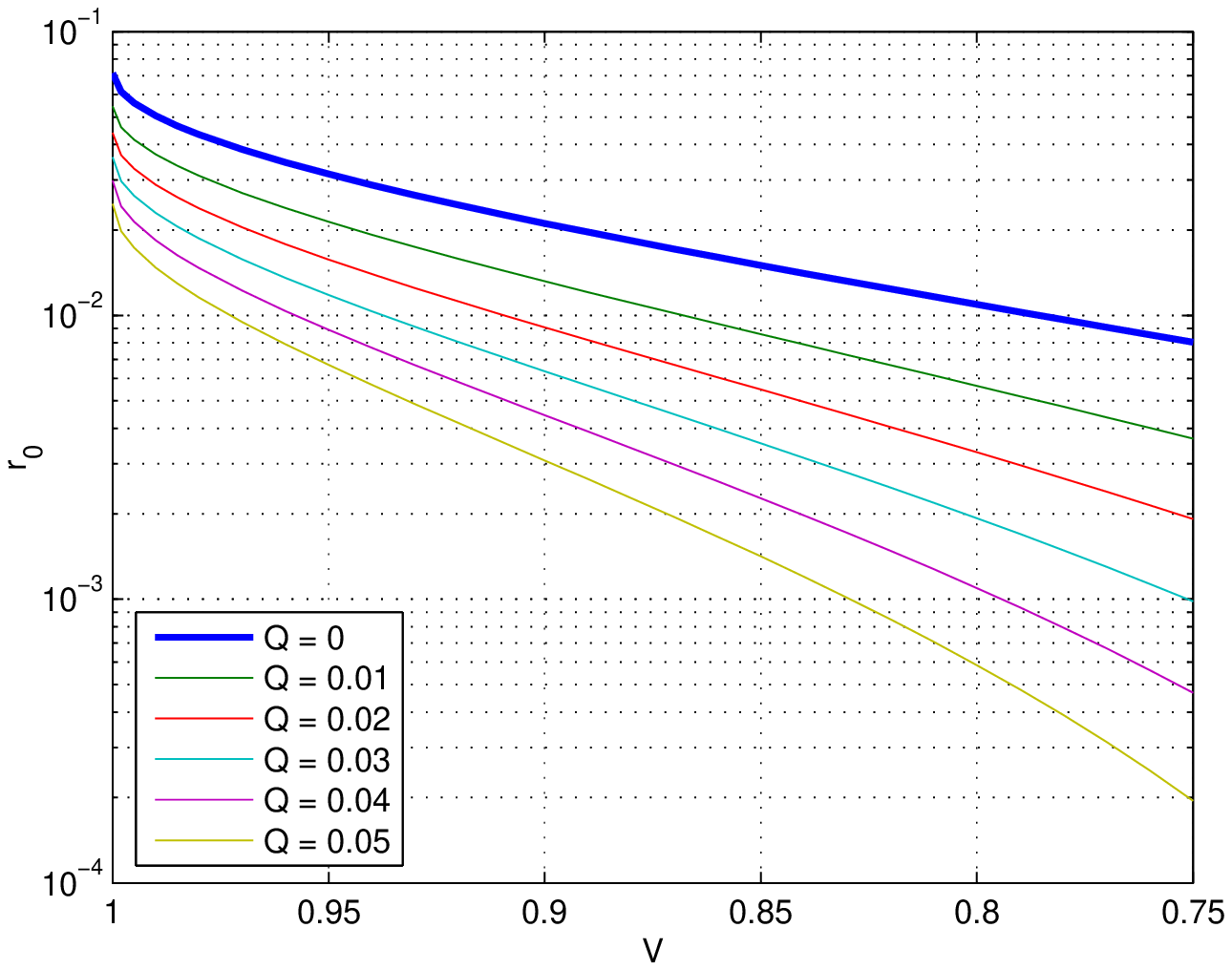} \caption{COWm2: $\mu_{opt}$ and $r_0$ for 1PA,
for different values of $Q$.} \label{fig_COW1}
\end{figure}
\end{center}

Having maximized Eve's information, one can run the one-parameter
optimization over the pulse intensity $\mu$. The optimal choice
$\mu_{opt}$ and the corresponding value of $r_0$ are plotted in
Fig.~\ref{fig_COW1}.

\section{Optimization of 1PA and 2PA on DPS}
\label{app_DPS_2}

As mentioned in the main text, the optimization of Eve's information
for a 2PA on DPS is more complicated than the one for COW, because
we could not find any evident simplification and had therefore to
start from the general formal expressions. For this reason, we find
it useful to sketch first the study of 1PA on DPS --- were it only
to show that our optimization on the 2PA yields indeed a more strict
bound.

\subsection{Optimization of 1PA on DPS}

\subsubsection{Eve's Attack and Information}

We have not studied 1PA on DPS in the main text, but the pattern is
the same as for all other attacks, so we just list the main points.
Eve's attacks is defined by (with $\sigma \in \{+,-\}$) \ba
\ket{\sigma \sqrt{\mu}}_A \ket{\cal{E}} &\longrightarrow & \ket{0}_B
\ket{v_\sigma}_E + \sigma \sqrt{\mu t} \ket{1}_B \ket{p_{\sigma}}_E
\ea so the unitarity condition and the visibility constraints read
\ba \braket{v_{+}}{v_{-}} = e^{-2\mu}&\equiv& \gamma
\label{con1dps1}\\ \forall \sigma, \omega \in \{+,-\}, \quad
\mbox{Re}\big[ \braket{v_\sigma}{p_\sigma}
\braket{p_\omega}{v_\omega}\big]&=&V\,. \ea This last condition
implies $\braket{v_+}{p_+} = \braket{v_-}{p_-} =
\sqrt{V}e^{i\tilde{\phi}}$, for some $\tilde{\phi}$ which won't play
any role, and which we set to 0; so we have \ba \braket{v_+}{p_+} =
\braket{v_-}{p_-} = \sqrt{V}\,. \label{con2dps1}\ea Note how all the
states participate in the constraints, contrary to what is the case
in all versions of COW.

To compute Eve's information, we start from Eve's conditioned states
$\rho_{E}^{A = \{\sigma \omega\}, B = b}
=\ket{\psi_{\sigma,\omega,b}}\bra{\psi_{\sigma,\omega,b}}$ with $
\ket{\psi_{\sigma,\omega,b}}=\sigma \ket{p_\sigma,v_\omega} + (-1)^b
\omega \ket{v_\sigma,p_\omega}$ and the mixtures $\rho_E^{a,b} =
\frac{1}{8}\sum_{\sigma} \rho_{E}^{A = \{\sigma\sigma(a)\}, B = b}$
where $\sigma(a)=(-1)^a\sigma$. Note that these states are not
normalized; rather, $\Tr\big(\rho_E^{a,b}\big)$ is equal to
$\frac{1+V}{2}=1-Q$ if $a=b$, and to $\frac{1-V}{2}=Q$ if $a\neq b$.
Finally, $\rho_E^{A = a} = \sum_{b}\rho_E^{a,b}$ and $\rho_E^{B = b}
= \sum_{b}\rho_E^{a,b}$ (now normalized). Eve's information is given
by $\chi_{AE}=\chi\left(\rho_E^{A = 0},\rho_E^{A = 1}\right)$, and
similarly for $\chi_{BE}$.

\subsubsection{Parametrization of Eve's states}

Eve's states can be chosen in a four-dimensional
Hilbert space. First, let's choose a basis with the first two vectors in the
subspace spanned by $\{\ket{v_+}, \ket{v_-}\}$, and in which
$\ket{v_+}$ and $\ket{v_-}$ read
\ba \ket{v_\sigma} &=& \left( \begin{array}{c} \sqrt{\frac{1+\gamma}{2}}
\\ \sigma\sqrt{\frac{1-\gamma}{2}} \\ 0 \\ 0
\end{array} \right)\ea so that (\ref{con1dps1}) is satisfied.
Let's also define $\ket{v_\sigma^\perp}$ as the orthogonal state to
$\ket{v_\sigma}$, in the subspace spanned by $\{\ket{v_+},
\ket{v_-}\}$: \ba \ket{v_\sigma^\perp} &=& \left( \begin{array}{c}
\sqrt{\frac{1-\gamma}{2}}
\\ -\sigma\sqrt{\frac{1+\gamma}{2}} \\ 0 \\ 0
\end{array} \right)\,. \ea The constraint (\ref{con2dps1}) on the visibility
implies that $\ket{p_\sigma}$ can be written as $\ket{p_\sigma} =
\sqrt{V} \ket{v_\sigma} - \sqrt{1-V} \ket{w_\sigma}$, where
$\ket{w_\sigma}$ is any (4-dimensional) state orthogonal to
$\ket{v_\sigma}$; this can be further decomposed as $\ket{w_\sigma}
= \cos\theta_\sigma e^{i \phi_\sigma} \ket{v_\sigma^\perp} +
\sin\theta_\sigma\ket{w_\sigma'}$ for some states
$\ket{w_{\sigma}'}$ orthogonal to both $\ket{v_+}$ and $\ket{v_-}$.
Finally we choose the last two vectors of the basis such that
$\ket{w_+'}$ and $\ket{w_-'}$ read
\ba \ket{w_+'} = \left( \begin{array}{c} 0 \\
0 \\ \cos(\theta/2)e^{i\phi/2}
\\ \sin(\theta/2)e^{i\phi/2}
\end{array} \right), \quad \ket{w_-'} = \left( \begin{array}{c} 0 \\
0 \\ \cos(\theta/2)e^{-i\phi/2}
\\ -\sin(\theta/2)e^{-i\phi/2}
\end{array} \right)\,. \ea In summary, for a given $V$ and a given $\mu$,
we are left without loss of generality with the six free parameters
$\theta_+, \theta_-, \theta, \phi_+, \phi_-, \phi$ that define \ba
\ket{p_\sigma} = \sqrt{V} \ket{v_\sigma} - \sqrt{1-V}
\cos\theta_\sigma e^{i \phi_\sigma} \ket{v_\sigma^\perp} -
\sqrt{1-V} \sin\theta_\sigma\ket{w_\sigma'}\,.\ea

\subsubsection{Results of the optimization}

The optimization over the six free parameters was performed
numerically. We find that Eve's optimal states have real
coefficients (the parameters $\phi_\pm, \phi$ can be chosen to be
0), and also that $\theta_+ = -\theta_-$. Once we fix this, there
remains only two free parameters to optimize.

Having maximized Eve's information, one can run the one-parameter
optimization over the pulse intensity $\mu$. The optimal choice
$\mu_{opt}$ and the corresponding value of $r_0$ are plotted in
Fig.~\ref{fig_DPS_1vs2}, along with the results for 2PA. In the case
$V=1$, this attack reduces to the BSA; in all other cases, the
optimal 1PA is manifestly less powerful than the best 2PA we have
found.

Note that after optimization, we find $\chi_{AE} \leq \chi_{BE}$:
Eve knows less about Alice's bit than about Bob's.

\subsection{Optimization of 2PA on DPS}

We now consider the 2PA on DPS, and we have to optimize $\chi_{AE}$
and $\chi_{BE}$ as given in (\ref{chidps}), submitted to the
constraints \ba \braket{v_{\sigma\omega}}{v_{\sigma\bar{\omega}}} =
\braket{v_{\sigma\omega}}{v_{\bar{\sigma}\omega}} =
e^{-2\mu}\,\equiv\,\gamma &\;,\;&
\braket{v_{\sigma\omega}}{v_{\bar{\sigma}\bar{\omega}}}\,=\,
e^{-4\mu} = \gamma^2\,, \label{con1dps2}\ea \ba \mbox{Re}\big[
\braket{p_{\sigma\omega}^{01}}{p_{\sigma\omega}^{10}}\big] =
\mbox{Re}\big[ \braket{v_{\sigma\omega}}{p_{\sigma\omega}^{01}}
\braket{p_{\sigma'\omega'}^{10}}{v_{\sigma'\omega'}}\big] &=& V
\label{con2dps2}\ea for all $\sigma,\omega,
\sigma',\omega'\in\{+,-\}$. We see that all the twelve states of Eve
appear in the expressions of the constraints.

\subsubsection{Parametrization of Eve's states}

Without any loss of generality, we choose the following symmetric
parametrization for Eve's four states $\ket{v_{\sigma \omega}}$: \ba
\ket{v_{++}} = \left( \begin{array}{c} \frac{1+\gamma}{2}
\\ \frac{1-\gamma}{2} \\ \sqrt{\frac{1-\gamma^2}{2}} \\ 0
\end{array} \right), \quad \ket{v_{--}} = \left( \begin{array}{c} \frac{1+\gamma}{2}
\\ \frac{1-\gamma}{2} \\ -\sqrt{\frac{1-\gamma^2}{2}} \\ 0
\end{array} \right), \quad \ket{v_{+-}} = \left( \begin{array}{c} \frac{1+\gamma}{2}
\\ -\frac{1-\gamma}{2} \\ 0 \\ \sqrt{\frac{1-\gamma^2}{2}}
\end{array} \right), \quad \ket{v_{-+}} = \left( \begin{array}{c} \frac{1+\gamma}{2}
\\ -\frac{1-\gamma}{2} \\ 0 \\ -\sqrt{\frac{1-\gamma^2}{2}}
\end{array} \right) \ea so that (\ref{con1dps2}) is satisfied. At
this point, we would have to optimize over Eve's most general states
$\ket{p_{\sigma\omega}^{10}}$ and $\ket{p_{\sigma\omega}^{01}}$ that
satisfied the constraints (\ref{con2dps2}). In general, these eight
states live in a 12-dimensional space, and the number of free
parameters is quite large.

In order to have a more tractable problem, we make some assumptions
(admittedly, we lose generality here). First, we look for states
that satisfy a more constraining version of (\ref{con2dps2}), namely
\ba \braket{p_{\sigma \omega}^{01}}{p_{\sigma \omega}^{10}} =
V&\;,\;& \braket{v_{\sigma \omega}}{p_{\sigma \omega}^{01}} =
\braket{v_{\sigma \omega}}{p_{\sigma \omega}^{10}} = \sqrt{V} \ea
for all $\sigma,\omega$. Then we can write \ba
\ket{p_{\sigma\omega}^{01}} = \sqrt{V} \ket{v_{\sigma\omega}} -
\sqrt{1-V} \ket{w_{\sigma\omega}^{01}} \\
\ket{p_{\sigma\omega}^{10}} = \sqrt{V} \ket{v_{\sigma\omega}} -
\sqrt{1-V} \ket{w_{\sigma\omega}^{10}} \ea with
$\braket{v_{\sigma\omega}}{w_{\sigma\omega}^{01}} =
\braket{v_{\sigma\omega}}{w_{\sigma\omega}^{10}} = 0$ and
$\braket{w_{\sigma\omega}^{01}}{w_{\sigma\omega}^{10}} = 0$. Note
that this may not be a true restriction: actually, for all the cases
treated above, an analog choice was found to be optimal. A more
serious restriction comes now: we suppose that the eight states
$\ket{w_{\sigma\omega}^{10}}$ and $\ket{w_{\sigma\omega}^{01}}$ live
in a 6-dimensional space and are parametrized only by real
coefficients. At this stage, we run the numerical optimization.

\subsubsection{Results of the optimization}

\begin{center}
\begin{figure}
\epsfxsize=7cm \epsfbox{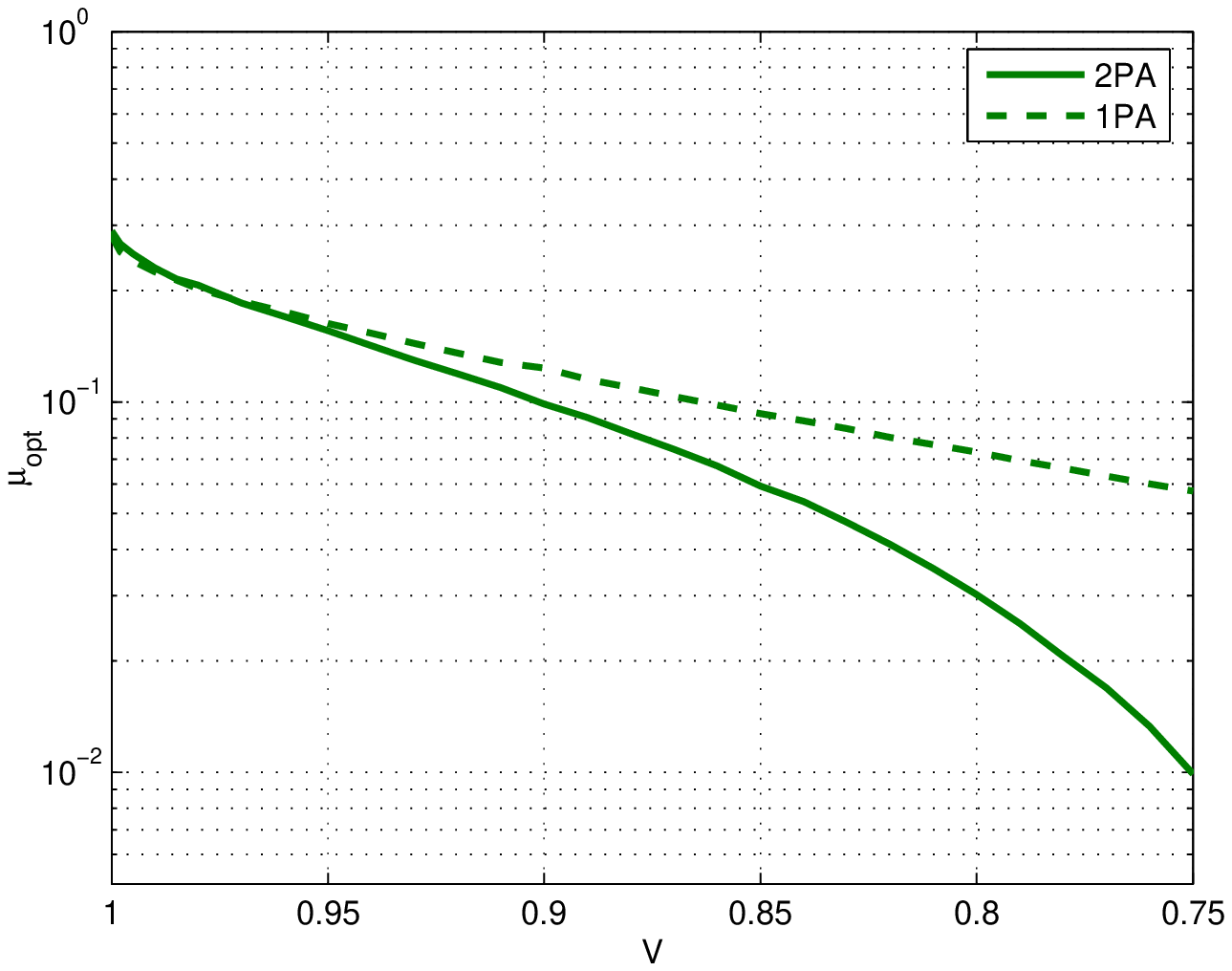} \epsfxsize=7cm
\epsfbox{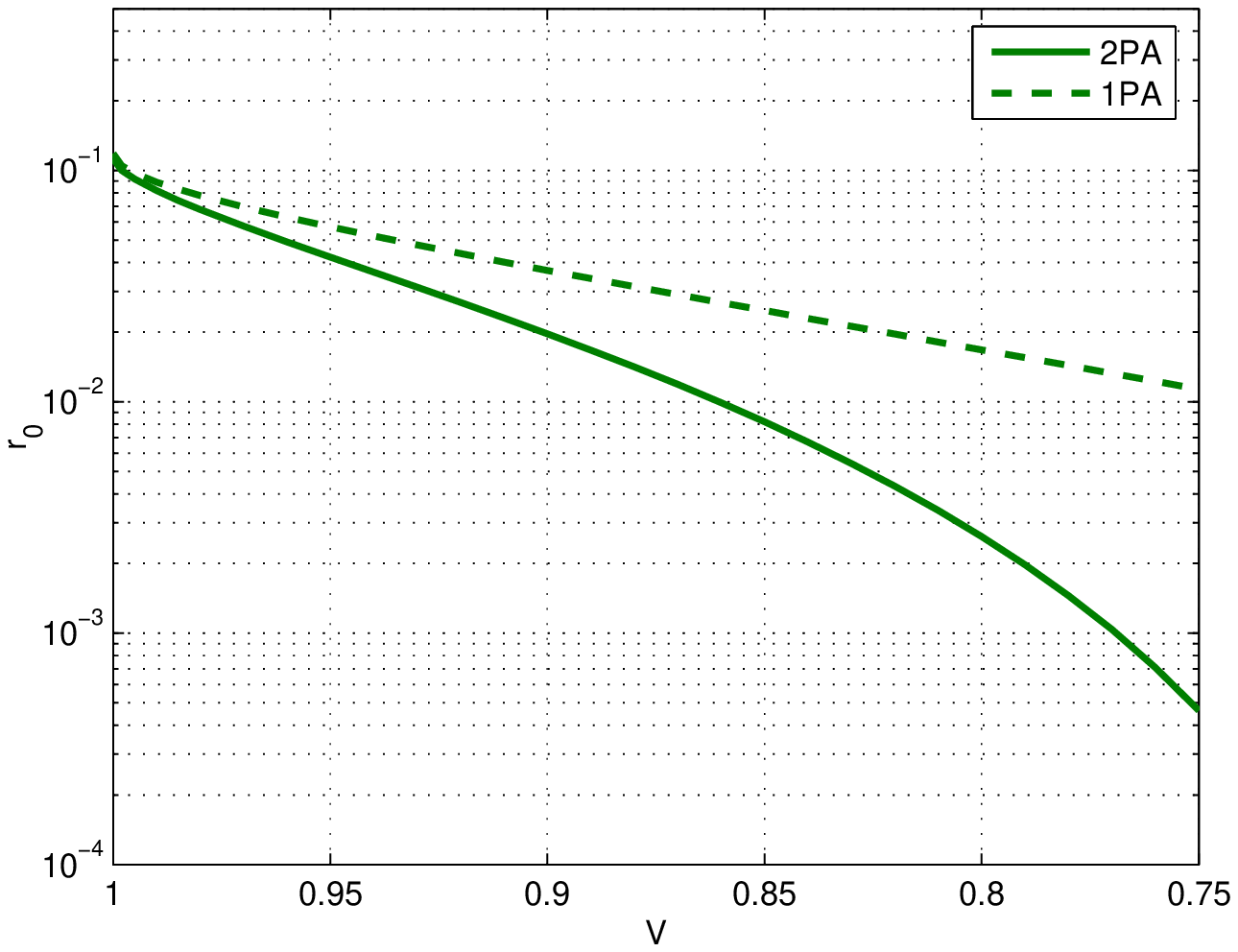} \caption{$\mu_{opt}$ and $r_0$ for 1PA
and 2PA on DPS.} \label{fig_DPS_1vs2}
\end{figure}
\end{center}

After having maximized Eve's information, one can run the
one-parameter optimization over the pulse intensity $\mu$. The
optimal choice $\mu_{opt}$ and the corresponding value of $r_0$ are
plotted in Fig.~\ref{fig_DPS_1vs2}, along with the results for 1PA.
In the case $V=1$, this attack reduces again to the BSA.

We didn't run the optimization aver all possible states, but we
believe that our results are very close to the optimal bounds we
could get for 2PA on DPS. Anyway, even though we might have missed
the true maximum of Eve's information, the attack we found and the
curves that are plotted still provide valid upper bounds, more
strict than the bounds given by 1PA on DPS.

Note finally that as for 1PA, we find after optimization $\chi_{AE}
\leq \chi_{BE}$: Eve knows less about Alice's bit than about Bob's.

\begin{thebibliography}{99}
\bibitem{footnote_attacks_zero_errors} Apart from BSA, other attacks exist that do not introduce errors
in distributed-phase-reference protocols: for instance,
photon-number-splitting attacks over the whole key, preserving the
coherence (these are hard to parametrize and have never been studied
in detail). For COW, there also exist zero-error attacks based on
unambiguous state discrimination \cite{bra07}.
\bibitem{footnote_distances} Of course, the distance cannot be too large: it has to be below
the distance where dark counts become important; and for COW, also
below the distance ($\sim 100$ km) at which other zero-error attacks
become possible \cite{bra07}.
\bibitem{footnote_steps_proof} Here are the main steps of the proof: first, one can easily show
that $\theta_0, \theta_1 \in [0,\pi/2]$, $\phi_0 = \phi_1 = 0$ and
$\braket{w_0}{w_1} = -1$ are optimal; once this is fixed, one can
show that for any given $\lambda$, the minimum of
$|\braket{p_{0\mu}^{01}}{p_{\mu0}^{10}}|$ over $\theta_0$ and
$\theta_1$ is obtained for $\theta_0 = \theta_1 = 0$; finally, one
can prove that $\lambda = 1$ is optimal.
\end{thebibliography}
\end{document}